\documentclass[9pt,twocolumn]{article}
\usepackage[mnras]{preprint}
\renewcommand*{\backrefalt}[4]{}

\usepackage{graphicx}	
\usepackage{amsmath}	

\usepackage{float}
\usepackage{xspace}
\usepackage{multirow}
\usepackage{float}

\newcommand{\myvec}[1]{\boldsymbol{#1}}
\newcommand{\myvecun}[1]{\hat{\boldsymbol{#1}}}
\newcommand{\mymatrix}[1]{\boldsymbol{\mathsf{#1}}}

\newcommand{\given}{\,|\,}

\newcommand{\mbh}{M_{\rm{BH}}}
\newcommand{\Dobs}{D_{\rm{obs}}}

\newcommand{\sgr}{\rm Sgr\ A ^\ast}

\newcommand{\Tp}{T_{\rm{p}}}

\newcommand{\SkytoBH}{
	\mathcal{R}_{\text{\raisebox{-1.ex}{\tiny sky$\to$BH}}}
}
\newcommand{\BHtoSky}{
	\mathcal{R}_{\text{\raisebox{-1.3ex}{\tiny BH$\to$sky}}}
}

\title[PPN--spin degeneracies in S62-like orbits]
{PPN--spin degeneracies in mock S62-like stellar-orbit inference}

\author[]{
	\Author{Shant Khlghatyan}{1}{0000-0001-7058-6156}
}

\date{
	\textsuperscript{1}A.I. Alikhanian National Science Laboratory, 2 Alikhanian Brothers Street, 0036 Yerevan, Armenia\\
}
\email{Sh.klghatyan@yerphi.am \\
	 Khlghatyanshant@gmail.com}

\begin{document}
\begin{journalinfo}[
	status=published,
	journal={Monthly Notices of the Royal Astronomical Society},
	shortjournal={MNRAS},
	doi={10.1093/mnras/stag1407},
	year={2026},
	volume={\textbf{551}},
	pages={1--22},
	
	firstfontsize=10,
	firstleading=12,
	
	runningfontsize=7,
	runningleading=8.5,
	]
\end{journalinfo}
	
	\twocolumn[
	\begin{@twocolumnfalse}
		
		\maketitle
		
		\begin{abstract}
			We investigate the degeneracies between black hole (BH) spin effects and parametrized post-Newtonian (PPN) parameters in the relativistic orbital dynamics of S2-like and S62-like stars orbiting Sagittarius A$^{\ast}$. Using a 1PN+SO Hamiltonian framework and synthetic astrometric and radial velocity datasets, we perform Bayesian parameter inference. For current baseline observational precisions, the dominant relativistic observable—the Schwarzschild periapsis advance—allows the recovery of the effective precession parameter $\Upsilon$, while leaving the individual PPN parameters $\gamma$ and $\beta$ degenerate. Assuming microarcsecond-level astrometric precision, the spin-induced Lense–Thirring signal becomes partially detectable; fixing the PPN sector to General Relativity allows the BH spin magnitude to be constrained to an uncertainty of $\sim10^{-2}$. However, simultaneously varying PPN and spin parameters reveals a strong, approximately linear covariance between $\Upsilon$ and the dimensionless spin parameter $\chi$. To overcome this limitation, we demonstrate that joint multi-star inference can disentangle the degeneracy by combining a wider-orbit star, which constrains the dominant 1PN sector, with a compact relativistic orbit that is sensitive to Lense–Thirring frame dragging.
		\end{abstract}
		
		
		\vspace{0.35cm}
		
	\end{@twocolumnfalse}
	]
	\printemailfootnote
	
	\section{Introduction}
	The Galactic center offers a unique natural laboratory for testing gravitational physics in the strong-field regime. At the dynamical center of the Milky Way lies the compact radio source $\sgr$, widely interpreted as a supermassive BH with a mass of about $4\times 10^6 M_{\odot}$. Surrounding $\sgr$ is a cluster of short-period stars withe different orbital parameters, commonly referred to as S-stars \citep{RevModPhys.82.3121}. These stars follow eccentric Keplerian-like orbits, but with measurable relativistic corrections. Their orbital dynamics are monitored  through astrometric measurements in the sky plane - right ascension and declination, as well as through spectroscopic observations that provide line-of-sight velocities. The combination of these observables allows for three-dimensional orbit reconstruction and enables stringent tests of General Relativity (GR).
	Among these objects, the star S2/S0--2 is of particular importance. It is one of the closest known stars to $\sgr$, with a semi-major axis of approximately $a \approx 0.123\,\arcsec$, a high eccentricity of $e \approx 0.88$, and a pericenter distance of about $120\,\mathrm{AU}$, corresponding to roughly $1400$ Schwarzschild radii \citep{gillessen2009monitoring}. Its orbital period of approximately $16$ years is sufficiently short that a full revolution has been observed with instruments such as the GRAVITY interferometer, which achieves astrometric precision at the $\sim10$ -- $100\,\upmu\mathrm{as}$ level \citep{abuter2021improved}. This temporal coverage is crucial, as it allows the measurement of relativistic effects accumulated over an entire radial period, including pericenter precession and gravitational redshift during pericenter passage.
	Observations around its 2018 pericenter passage provided the first direct detection of the combined gravitational redshift and transverse Doppler effect in the Galactic center, consistent with GR predictions. Continued high-precision monitoring has enabled increasingly stringent constraints on the Schwarzschild pericenter precession, measured to be approximately $12 '$ per orbital period \citep{abuter2020detection}. In addition to S2, several other short-period stars have been identified in the Galactic center. In particular, S62 has an orbital period of approximately $9.9\,\mathrm{yr}$ and exhibits a higher eccentricity than S2 \citep{peissker2020s62}. Similarly, S301 has an orbital period of about $8.7\,\mathrm{yr}$ \citep{abdeld2026discovery}, while S4716 completes an orbit in only $\sim 4\,\mathrm{yr}$ \citep{peissker2022observation}. Owing to their short orbital periods and close pericenter passages, these stars may provide valuable constraints on the spin of $\sgr$, which remains an active subject of investigation \citep{abdeld2026discovery}.
	
	While early stellar-orbit constraints gave an upper limit $\chi \lesssim 0.1$ \citep{fragione2020upper}, more recent analyses combining X-ray and radio observations via the "outflow method" yield $\chi = 0.90 \pm 0.06$ \citep{daly2024new}. The Event Horizon Telescope imaging of the compact source also favors a rapidly rotating BH, with spin estimates around $\chi \approx 0.9$ \citep{andrianov2024estimation,janssen2025deep} ($\chi\in[0,1]$ is the dimensionless spin parameter). This high spin would induce Lense–Thirring precession (LT) in the S-star orbits, offering a new test of GR in the strong-field regime. Detecting BH spin via stellar orbits requires stars on sufficiently close, high-precision monitored orbits, since spin-induced effects such as LT precession are much smaller than the leading-order Schwarzschild effects and are easily masked by Newtonian perturbations from other stars. As shown in \citealt{waisberg2018stellar}, only stars with very tight orbits (typically within a few hundred Schwarzschild radii) and  astrometric monitoring at the level of $\sim 10\, \upmu\rm{as}$ are capable of producing a detectable spin signature in the Galactic Centre environment. 
	
	The interpretation of the observational data relies on a range of theoretical models. These include relativistic treatments within GR, as well as alternative gravity scenarios including scalar–tensor theories and Yukawa-type modifications of the gravitational interaction \citep{grould2017general,borka2013constraining,bambhaniya2024relativistic,navarrete2026testing}. In addition, models incorporating an extended mass distribution, such as a dark matter cusp \citep{heissel2022dark,abd2024improving}, can produce measurable deviations from purely Keplerian motion and may potentially mimic or alter relativistic signatures.
	
	In these studies, the post-Newtonian (PN) approximation is commonly employed, often within the PPN framework \citep{Will:2018bme,poisson2014gravity}. The PPN formalism provides a systematic way to go beyond GR by introducing a set of parameters that characterize deviations from GR in the metric perturbations. These PPN parameters allow one to test and constrain alternative theories of gravity.
	
	The PPN formulation depends on the choice of gauge. Commonly used gauges include the harmonic gauge and the Arnowitt--Deser--Misner (ADM) gauge \citep{schafer2024hamiltonian}. Starting from the second post-Newtonian order, the Lagrangian in the harmonic gauge contains acceleration-dependent terms, in contrast to the ADM gauge. Nevertheless, both formulations describe the same physical dynamics and are related by suitable coordinate transformations \citep{blanchet2014gravitational}.
	
	Within the PN/PPN framework, the two-body problem admits a quasi-Keplerian parametrization of the orbital motion. Such parametrized solutions are known up to first PN (1PN) order \citep{damour1985general}, extended to third PN order \citep{memmesheimer2004third}, and including spin--orbit (SO) coupling terms, which formally enter at 1.5PN order \citep{konigsdorffer2005post}. These quasi-Keplerian parametrizations are particularly useful for rapid parameter inference, for example in the context of S-stars, although their extension to include additional perturbative effects, such as an extended mass distribution, is generally non-trivial \citep{hyman2023analytic}.
	
	The number of independent PPN parameters that contribute depends on the perturbative order of the expansion; in general, ten parameters are required \citep{Will:2018bme}. Higher post-Newtonian orders probe increasingly complex components and nonlinear structures of the metric, thereby introducing additional PPN parameters. At 1PN order, including SO corrections, only two PPN parameters arise: $\gamma$ and $\beta$, with general relativity corresponding to $\gamma=\beta=1$. The parameter $\gamma$ governs the linear spatial curvature generated per unit mass and enters the spatial components $g_{ij}$ of the metric at leading PN order, while $\beta$ controls the nonlinear self-interaction of the gravitational potential and appears at quadratic order in the time--time component $g_{00}$. Although in the standard PPN gauge these parameters are cleanly separated between metric sectors, different coordinate choices or explicit solution expansions can redistribute PN corrections between components, leading to apparent mixing of $\gamma$-dependent contributions into $g_{00}$ without altering their physical interpretation.
	
	Current Solar System experiments constrain deviations to $|\gamma-1|\lesssim 2.3\times10^{-5}$ \citep{bertotti2003test} and $|\beta-1|\lesssim 8\times10^{-5}$ \citep{genova2018solar}. The motion of S-stars near the Galactic center provides an independent probe of these parameters in a much stronger gravitational regime. At the 1PN order, the Schwarzschild periapsis precession depends only on a specific linear combination of $\gamma$ and $\beta$, leading to a degeneracy when using orbital data alone. However, as shown in \citep{de2025future}, this degeneracy can be broken by combining orbital measurements of the S2 star with gravitational lensing (GL) observations of stars such as S62, since lensing effects depend solely on $\gamma$. This joint analysis enables independent constraints on both parameters.
	
	In this work, we consider the 1PN+SO system in the test-particle approximation for the $\sgr$ BH environment. We generate synthetic astrometric and radial velocity (RV) observations for S2- and S62-like stellar orbits with a precision of $\sim 80,\upmu\mathrm{as}$, together with enhanced S62-like datasets with higher astrometric accuracy. Using these datasets, we perform Markov chain Monte Carlo (MCMC) parameter inference to investigate the recoverability of the BH spin signal and its degeneracies with the PPN parameters.
	
	The remainder of this paper is organized as follows: in Section~\ref{sec:Models} we describe the orbital model and the procedure used to generate mock observations with the adopted uncertainty model; in Section~\ref{sec:Datasets} we present the datasets; and in Section~\ref{seq:Inference} we discuss parameter-space degeneracies and correlations, and describe the maximum a posteriori (MAP) and MCMC inference procedures. In Section~\ref{seq:results}, we present and discuss the results obtained from the MCMC analysis.

	\section{Physical model}\label{sec:Models}
	In this section, we present the orbital, astrometric, and spectroscopic models employed in our analysis. The orbital dynamics are described by the 1PN$+$SO Hamiltonian in the ADM gauge, which has the following form:
	\begin{align}\label{eq:ADMhamiltonian}
		H(\myvec{R},\myvec{P})=&\frac{\myvec{P}^2}{2\mu}-\frac{GM\mu}{R}\\
		&+\frac{1}{c^2}\Bigg\{\frac{1}{8}(3\eta-1)\frac{\myvec{P}^4}{\mu^3}-\frac{1}{2}\frac{GM\mu}{R}\Bigg[(3+\eta)\frac{\myvec{P}^2}{\mu^2}\nonumber \\
		&\,\,+\frac{1}{M\mu}\frac{1}{R^2} (\myvec{R}\cdot\myvec{P})^2\Bigg]+\frac{G^2M^2\mu}{2R^2}\Bigg\}\nonumber\\
		&+\frac{G}{c^2R^3}(\myvec{R}\times\myvec{P})\cdot\myvec{S}_{\rm eff},\nonumber\\
	\end{align}
	where $M$ is the total mass of two-body system, $\mu=m_1m_2/M$ is reduced mass, $\eta=\mu/M$, $\myvec{R}$ relative separation vector and $\myvec{P}$ is the conjugated momentum vector. The quantity $\myvec{S}_{\rm eff} = \left(2+{3m_2}/{2m_1}\right)\myvec{S}_1 + \left(2+{3m_1}/{2m_2}\right)\myvec{S}_2$ denotes the effective spin.
	In the presence of the PPN parameters $\gamma$ and $\beta$, the Hamiltonian takes the form (see, for example, \citealt{barker1976lagrangian})
	\begin{align}\label{eq:ADMPPNhamiltonian}
		H(\myvec{R},\myvec{P})=&\frac{\myvec{P}^2}{2\mu}-\frac{GM\mu}{R}\\
		&+\frac{1}{c^2}\Bigg\{\frac{1}{8}(3\eta-1)\frac{\myvec{P}^4}{\mu^3}-\frac{1}{2}\frac{GM\mu}{R}\Bigg[(1+2\gamma+\eta)\frac{\myvec{P}^2}{\mu^2}\nonumber \\
		&\,\,+\frac{1}{M\mu}\frac{1}{R^2} (\myvec{R}\cdot\myvec{P})^2\Bigg]+(2\beta-1)\frac{G^2M^2\mu}{2R^2}\Bigg\}\nonumber\\
		&+\frac{G}{c^2R^3}(\myvec{R}\times\myvec{P})\cdot\tilde{ \myvec{S}}_{\rm eff}\nonumber
	\end{align}
	where the effective spin vector is modified according to $\tilde{\myvec{S}}_{\rm eff}=\left[\gamma+1+\left(\gamma+{1}/{2}\right){m_2}/{m_1}\right]\myvec{S}_1+\left[\gamma+1+\left(\gamma+{1}/{2}\right){m_1}/{m_2}\right]\myvec{S}_2$
	
	\subsection{Orbital model}
	We consider the two-body problem in the test-particle approximation (i.e.\ the Kepler problem), consisting of a central BH such that $m_1 \equiv \mbh \gg m_2$ and $\myvec{S}_2 \equiv 0$. We additionally assume a constant BH spin vector, $\myvec{S}_1 = G\mbh^2 c^{-1}\chi \myvecun{s}$, where $\myvecun{s}$ is a constant unit vector and $0 \leq \chi \leq 1$ is the dimensionless spin parameter. Under these assumptions, the Hamiltonian in equation~\eqref{eq:ADMPPNhamiltonian} reduces to the form
	\begin{align}\label{eq:ADMmyhamiltonian}
		\mathcal{H}(\myvec{r},\myvec{p}) = & \frac{p^2}{2} - \frac{G \mbh }{r} \\ &+\frac{1}{c^2} \left[ -\frac{p^4}{8} - \frac{G \mbh}{2 r} (1 + 2 \gamma) p^2 + \frac{G^2 \mbh^2 }{2 r^2} (2 \beta - 1) \right]\nonumber\\
		&+ \frac{1}{c^3} \frac{G^2 \mbh^2}{r^3}\chi(\gamma + 1) \myvecun{s} \cdot \myvec{L}\nonumber
	\end{align}
	where $r:=|\myvec{r}|$ is the test particle radius relative to BH, $p^2:=\myvec{p}\cdot\myvec{p},\, \myvec{L}:=\myvec{r} \times \myvec{p}$ is the canonical angular momentum. There are three conserved quantities $(\mathcal{H}, \myvec{L}\cdot\myvecun{s}, |\myvec{L}|)$.
	Corresponding Hamilton equations are
	\begin{align}\label{eq:HamiltonEq}
		\dot{\myvec{r}}&=\frac{\partial \mathcal{H}(\myvec{r},\myvec{p})}{\partial\myvec{p}}\\
		&=\myvec{p}-\frac{1}{c^2}\left[\frac{p^2}{2}+\frac{G\mbh}{r}(1+2\gamma)\right]\myvec{p}+\frac{1}{c^3}\frac{G^2\mbh^2}{r^3}\chi (\gamma+1)\myvecun{s}\times\myvec{r}, \nonumber \\ 
		\dot{\myvec{p}}&=-\frac{\partial\mathcal{H}(\myvec{r},\myvec{p})}{\partial\myvec{r}} \\ 
		&=-\frac{G \mbh}{r^3}\myvec{r}
		+\frac{1}{c^2}\frac{G\mbh}{r^3}
		\left[ \frac{G \mbh}{r}(2\beta-1)
		-\frac{p^2}{2}(1+2\gamma)
		\right]\myvec{r} \nonumber \\
		&\quad
		+\frac{1}{c^3} \frac{G^2 \mbh^2}{r^3} \chi (\gamma+1)
		\left[
		3\frac{\myvec{r}}{r^2}(\myvecun{s}\cdot \myvec{L}) 
		- \myvec{p} \times \myvecun{s}
		\right]\nonumber
	\end{align}
	In many applications, it is convenient to express the equations of motion in terms of the coordinate acceleration rather than the canonical momentum. This form can be obtained by inverting the relation $\dot{\myvec{r}}(\myvec{r}, \myvec{p})$ perturbatively up to the required PN order, thereby expressing $\myvec{p}$ as a function of $\myvec{r}$ and $\dot{\myvec{r}}$. Substituting this expression into Hamilton equation for $\dot{\myvec{p}}(\myvec{r}, \myvec{p})$ yields an equation of motion of the form $\ddot{\myvec{r}}(\myvec{r}, \dot{\myvec{r}})$. 
	
	The relativistic dynamics of a test particle orbiting a BH induce both a periapsis advance and a precession of the orbital plane, including secular variations of the longitude of the ascending node and the orbital inclination. These secular effects are critical for testing GR in extreme environments, such as through tracking the trajectories of S-stars orbiting close to $\sgr$. The total secular periapsis shift per orbit can be decomposed into the non-spinning, spherically symmetric 1PN Schwarzschild contribution and the 1.5PN Lense--Thirring contribution induced by the BH spin: $\Delta\omega_{\rm tot}=\Delta\omega_{\rm{1PN}}+\Delta\omega_{\rm{1.5PN}}$. The Schwarzschild contribution, parameterized by the PPN parameters $\gamma$ and $\beta$, yields a purely in-plane periapsis advance per orbit:
	\begin{equation}\label{eq:SchPrec}
		\Delta\omega_{\rm{1PN}}= 6\pi\epsilon\Upsilon,
	\end{equation}
	where  $\Upsilon := ({2 + 2\gamma - \beta})/{3}$ is effective Schwarzschild precession parameter and $\epsilon := {G \mbh}/{a_{\rm sma}(1-e^2)c^2}$ is PN strength parameter. Here $a_{\rm sma}$ is the semi-major axis and $e$ is the orbital eccentricity. In contrast, the LT effect at 1.5PN order breaks spherical symmetry, inducing both a nodal precession (precession of the orbital plane) and an additional contribution to the argument of periapsis. For a central BH with dimensionless spin parameter $\chi$ and spin-direction unit vector $\myvecun{s}$, the nodal shift per orbit is given by
	\begin{equation}\label{eq:LTOmegaPrec}
		\Delta\Omega_{\rm{1.5PN}} = 2\pi \epsilon^{\frac{3}{2}}\chi (\gamma + 1) (\myvecun{s}\cdot\myvecun{m})\csc\iota
	\end{equation}
	while the corresponding LT contribution to the argument of periapsis is
	\begin{equation}\label{eq:LTomegaPrec}
		\Delta\omega_{\rm{1.5PN}} = -2\pi\epsilon^{\frac{3}{2}}\chi (\gamma + 1)\left[2(\myvecun{s}\cdot\myvecun{l})+(\myvecun{s}\cdot\myvecun{m})\cot\iota\right],
	\end{equation}
	the final term arises from the coupling between the nodal precession and the orbital geometry. Here, $\iota$ is the orbital inclination, and $(\myvecun{n}, \myvecun{m}, \myvecun{l})$ form an orthonormal triad defined by the line of nodes: $\myvecun{n}$ is the unit vector along the ascending node, $\myvecun{l}$ is the unit vector normal to the orbital plane, and $\myvecun{m} = \myvecun{l} \times \myvecun{n}$ (for more details see Section~\ref{sec:app:precession} and Fig.~\ref{fig:RefFrames}). For higher-order (quadrupole) contributions to the orbital precession, see \citep{refId0}.

	\subsection{Reference frames}\label{sec:refframe}
\begin{figure}
		\includegraphics[width=\columnwidth]{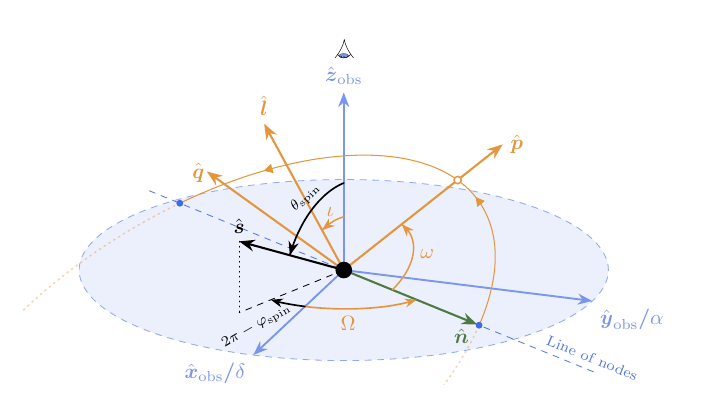}
		\caption{Schematic illustration of the reference frames used in this work. The observer frame (\textit{blue}) is defined by $(\myvecun{x}_{\rm obs}, \myvecun{y}_{\rm obs}, \myvecun{z}_{\rm obs})$, where 
			$\myvecun{z}_{\rm obs}$ points toward the observer located at a distance $\Dobs$. 
			The instantaneous orbital frame (\textit{orange}) is specified by the longitude 
			of the ascending node $\Omega$, the argument of periapsis $\omega$, and the inclination 
			$\iota$. The \textit{black vector} denotes the BH spin unit vector $\myvecun{s}$, shown 
			with its orientation relative to the observer frame.}
		\label{fig:RefFrames}
	\end{figure}

	In this work, we employ three reference frames, schematically illustrated in Fig.~\ref{fig:RefFrames}:
	\begin{itemize}
		\item Observer frame.
		We define a right-handed Cartesian frame $(\myvecun{x}_{\rm obs}, \myvecun{y}_{\rm obs}, \myvecun{z}_{\rm obs})$ centered on the BH. The $\myvecun{z}_{\rm obs}$-axis points from the BH toward the observer. The two orthonormal vectors $(\myvecun{x}_{\rm obs}, \myvecun{y}_{\rm obs})$ span the plane of the sky: $\myvecun{x}_{\rm obs}$ (Declination), and $\myvecun{y}_{\rm obs}$ (Right Ascension). This frame is used to compute observable quantities such as projected sky-plane coordinates, photon arrival times, and spectroscopic radial velocities.
		We note that alternative conventions are also used in the literature, in which the observer-frame $z$-axis is defined in the opposite direction and the inclination is correspondingly defined with the opposite sign, $i = -\iota$ (see, e.g., Fig.~1 and Fig.~C1 of~\citealt{heissel2022dark}).
		\item BH frame.
		The BH frame is an inertial frame in which the equations of motion are integrated. Its $z$-axis, $\myvecun{z}_{\rm BH}$, is aligned with the BH unit spin vector, $\myvecun{s}$. In the observer frame, this spin direction is parameterized by the polar and azimuthal angles $\theta_{\rm spin}$ and $\varphi_{\rm spin}$. The remaining orthogonal axes, $\myvecun{x}_{\rm BH}$ and $\myvecun{y}_{\rm BH}$, are chosen arbitrarily within the equatorial plane of the BH to complete a right-handed Cartesian triad.
		
		\item Perifocal (orbital) frame.
		The perifocal frame defines the orientation of the orbit in space through the Keplerian orbital elements $(\iota, \Omega, \omega)$. It is spanned by the orthonormal basis $(\myvecun{p}, \myvecun{q}, \myvecun{l})$, where $\myvecun{p}$ points toward periapsis, $\myvecun{q}$ is rotated by $90^\circ$ in the direction of motion within the orbital plane, and $\myvecun{l} = \myvecun{p} \times \myvecun{q}$.
		In the observer frame, these basis vectors are obtained via the standard Thiele–Innes construction. They define the orbital plane and are used to construct the instantaneous state vectors. The explicit expressions and the procedure for constructing the initial conditions, as well as the transformation to the BH frame used for numerical integration, are given in Section~\ref{subsec:app:TransformIC}.
		
	\end{itemize}

	\subsection{Observables \& time delays}\label{subsec:obsevables}
	The computation of observables consists of two steps: (i) geometrical projection of the orbit onto the observer frame, and (ii) mapping of the emission time to the observed time through relativistic time delays. We assume a static observer at infinity.
	After integrating the equations of motion in the BH frame, the trajectory is transformed into the observer frame (see Section~\ref{subsec:app:TransformIC}), yielding $(\myvec{r}_{\rm obs}, \myvec{v}_{\rm obs})$. The angular position on the sky is then given by the small-angle approximation
	\begin{equation}
		\Delta\alpha = \frac{y_{\rm obs}}{\Dobs}, \qquad
		\Delta\delta = \frac{x_{\rm obs}}{\Dobs},
	\end{equation}
	where $\Dobs$ is the distance between the observer and the BH. In this convention, the BH is placed at the origin of the observer sky plane, such that $(\Delta\alpha,\Delta\delta)=(0,0)$.
	In addition to the geometrical projection, we include the leading astrometric GL correction through the primary-image weak-lensing equation,
	\begin{equation}\label{eq:GLastro}
		\Delta\alpha_{\rm GL} = \Delta\alpha + \updelta\alpha_{\rm GL},
		\qquad
		\Delta\delta_{\rm GL} = \Delta\delta + \updelta\delta_{\rm GL},
	\end{equation}
	where $\Delta\alpha_{\rm GL}$ and $\Delta\delta_{\rm GL}$ denote the lensed apparent astrometric position, while $\updelta\alpha_{\rm GL}$ and $\updelta\delta_{\rm GL}$ are the corresponding GL displacements (for more details about GL model see Section~\ref{sec:app:GL}).
	
	The observed time is related to the emission time via
	\begin{equation}\label{eq:obstime}
		t_{\rm obs}(t_{\rm em}) = t_{\rm em} + \Delta_{\rm R}(t_{\rm em}) + \Delta_{\rm Sh}(t_{\rm em}),
	\end{equation}
	where we include the R\o mer delay (geometric light-travel time) and the Shapiro delay:
	\begin{align}\label{eq:Delays}
		\Delta_{\rm R}(t_{\rm em}) &= -\frac{ \myvecun{z}_{\rm{obs}} \cdot \myvec{r}(t_{\rm em})}{c}, \\
		\Delta_{\rm Sh}(t_{\rm em}) &= -(1+\gamma)\frac{G M_{\rm BH}}{c^3}
		\ln\!\left[\frac{r(t_{\rm{em}}) + \myvecun{z}_{\rm{obs}} \cdot \myvec{r}(t_{\rm em})}{2\Dobs}\right].
	\end{align}
	Finally, the total time delay is defined up to an additive constant, which is fixed by subtracting its value at the reference epoch, which is chosen as periapsis passage epoch $\Tp$. For a spinning BH, the time delay also receives a contribution from the gravitomagnetic LT effect \citep{ciufolini2003gravitomagnetic}. However, this contribution is suppressed relative to the Shapiro delay by an additional factor of $v/c$, and is expected to be at the millisecond level for S-star orbits around $\sgr$. We therefore neglect it in the present analysis.
	During inference, we require the inverse mapping $t_{\rm em}(t_{\rm obs})$, which can be obtained in several ways; for details, see Section~\ref{subsec:app:timemapping}.
	
	We express the observed redshift in velocity units by defining $v_{\rm rad} \equiv cz$, where $z$ denotes the total relativistic redshift.
	\begin{align}\label{eq:RV}
		v_{\rm rad} =
		&-\myvec{v} \cdot \myvecun{z}_{\rm{obs}}
		+ \frac{v^2}{2c}
		+ \frac{G \mbh}{r c}  \\
		&- (1+\gamma)\frac{G\mbh}{c^2} \frac{1}{r +\myvecun{z}_{\rm{obs}} \cdot \myvec{r}} \left( \myvec{v}\cdot\myvecun{z}_{\rm{obs}}  + \frac{(\myvec{r} \cdot \myvec{v})}{r} \right).\nonumber
	\end{align}
	The first term represents the classical line-of-sight velocity. The second and third terms correspond to the transverse Doppler effect and gravitational redshift, respectively, while the last term arises from the Shapiro time delay. Hereafter, the corresponding Shapiro contribution is denoted by $v_{\rm rad}^{\rm Sh}$.
	
	Also, radial velocity measurements may be available at epochs where no astrometric data are recorded, and vice versa. For realism, we consider observational data consisting of astrometric and spectroscopic mock measurements, $\{(\Delta \alpha(t_i), \Delta \delta(t_i))\}_{i=1}^{N_{\rm pos}}, \, \{v_{\rm rad}(t_j)\}_{j=1}^{N_{\rm RV}}$,
	where, in general, the time stamps $\{t_i\}$ and $\{t_j\}$ do not coincide. That is, astrometric and radial velocity measurements are obtained at different epochs, and the model predictions are evaluated independently at the corresponding observation times for each dataset.
	
	The mock observation epochs are selected from a densely sampled reference orbit. To mimic the seasonal visibility of the Galactic Centre, we first define a set of visible epochs by retaining only points that fall within a prescribed annual observing window; in the mock datasets used here, this window is taken to run from March to October. The total numbers of astrometric and radial velocity measurements, $N_{\rm pos}$ and $N_{\rm RV}$, are then set as fixed fractions of this visible set. Astrometric and radial velocity epochs are sampled independently. In the default sampling mode, the observation epochs are drawn uniformly at random without replacement from the visible epochs. We also allow for enhanced sampling around periapsis. In this case, specified fractions $f_{\rm peri}^{\rm ast}$ and $f_{\rm peri}^{\rm RV}$ of the astrometric and radial-velocity samples are forced to lie near periapsis. The periapsis passages are taken to occur at $t_{{\rm p},k}=T_{\rm p}+kT_{\rm 1PN}$, where $T_{\rm 1PN}$ is the 1PN radial period. For each periapsis passage covered by the simulated time span, candidate epochs are defined by $|t-t_{{\rm p},k}|\leq \Delta t_{\rm p}$. The required periapsis samples are distributed approximately evenly among these periapsis passages and are drawn uniformly at random, without replacement, from the candidate epochs in each periapsis window. The remaining observations are drawn uniformly at random from the seasonally visible epochs. Thus, the periapsis prescription guarantees an enhanced number of measurements close to periapsis, while additional randomly selected epochs may also fall in the same time intervals by chance.

	We add Gaussian noise to both the astrometric and radial velocity observables. For astrometry, the measured angular offsets are modeled as
	\begin{equation}
		\begin{pmatrix}
			\Delta \alpha_{\rm obs}(t_i) \\
			\Delta \delta_{\rm obs}(t_i)
		\end{pmatrix}
		=
		\begin{pmatrix}
			\Delta \alpha(t_i) \\
			\Delta \delta(t_i)
		\end{pmatrix}
		+
		\myvec{\epsilon}_{{\rm ast},i},
	\end{equation}
	where $\boldsymbol{\epsilon}_{\rm ast}$ is a zero-mean Gaussian random vector with covariance matrix
	\begin{equation}\label{eq:adeltacov}
		\mymatrix{C}_i =
		\begin{pmatrix}
			\sigma_{\alpha,i}^2 & \rho_i\,\sigma_{\alpha,i}\sigma_{\delta,i} \\
			\rho_i\,\sigma_{\alpha,i}\sigma_{\delta,i} & \sigma_{\delta,i}^2
		\end{pmatrix}.
	\end{equation}
	The astrometric uncertainties $(\sigma_{\alpha,i}, \sigma_{\delta,i})$ are assumed to follow log-normal distributions whose parameters depend on the observational precision tier. The correlation coefficient $\rho_i$ is drawn from a uniform distribution within a tier-dependent range.
	Similarly, the observed radial velocity is modeled as
	\begin{equation}
		v_{\rm rad,obs}(t_j) = v_{\rm rad}(t_j) + \epsilon_{{\rm rv},j}, 
		\qquad 
		\epsilon_{{\rm RV},j} \sim \mathcal{N}(0, \sigma_{{\rm RV},j}^2),
	\end{equation}
	where the radial velocity uncertainties $\sigma_{{\rm RV},j}$ are also drawn from log-normal distributions depending on the spectroscopic precision tier.
	
	For each observation, the corresponding precision tier is assigned probabilistically, allowing for a mixture of different observational regimes within a single dataset. This approach enables a flexible and realistic modeling of heterogeneous observational campaigns, combining standard-resolution measurements with high-precision data.
	
	The adopted distributions for the noise amplitudes and correlation coefficients are summarized in Table~\ref{tab:noise_model}. These correspond to typical uncertainties of $\sim 1\,{\rm mas}$ for the standard regime and $\sim 80\,\upmu{\rm as}$ for the high-precision regime.
	
	\begin{table}
		\centering
		\caption{
			Noise model for mock observations. Astrometric uncertainties are drawn from log-normal distributions such that 
			$\ln \sigma_{\alpha,\delta} \sim \mathcal{N}(\ln \mu_{\mathrm{ast}}, s^2)$, where $\mu_{\mathrm{ast}}$ denotes the median uncertainty (in arcsec) and $s$ is the log-space scatter. 
			The correlation coefficient $\rho$ is drawn from a uniform distribution within the specified range. 
			Radial velocity uncertainties are also modeled as log-normal distributions with median $\mu_{\mathrm{RV}}$ and log-scatter $s$.
		}
		\label{tab:noise_model}
		\begin{tabular}{l|ccc|cc}
			\hline
			\hline
			& \multicolumn{3}{c|}{Astrometry} & \multicolumn{2}{c}{RV} \\
			Tier & $\mu_{\mathrm{ast}}$ & $s$ & $\rho$ range & $\mu_{\mathrm{RV}}\,$ & $s$ \\
			\hline
			Low  & $1\,[\rm{mas}]$        & 0.4 & $[-0.4,\,0.4]$ & $30\,  [\mathrm{km\ s^{-1}}]$ & 0.3 \\
			High & $80\, [\upmu\rm{as}]$ & 0.2 & $[-0.2,\,0.2]$ & $5\,  [\mathrm{km\ s^{-1}}]$  & 0.2 \\
			\hline
		\end{tabular}
	\end{table}
	
	The fraction of high- and low-precision measurements in the dataset is governed by the precision fraction parameters $f_{\rm High}^{\rm ast}$ and $f_{\rm High}^{\rm RV}$ for astrometric and radial velocity data, respectively. For example, $f_{\rm High}^{\rm ast} = 1$ corresponds to a dataset composed entirely of high-precision astrometric observations.
	
	\section{Synthetic datasets}\label{sec:Datasets}
\begin{figure}
		\centering
		\includegraphics[width=\columnwidth]{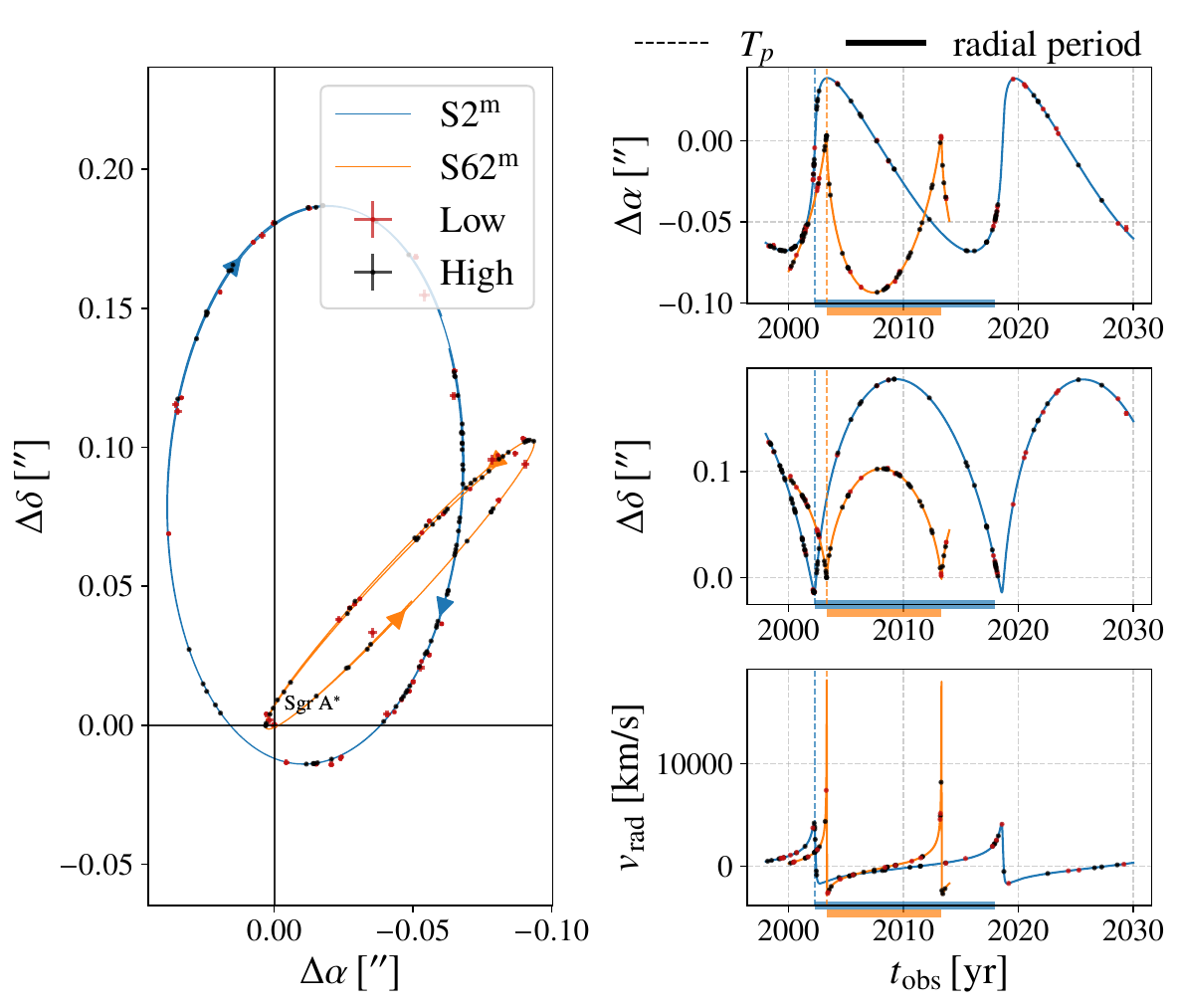}
		\caption{Synthetic astrometric and radial velocity data for the mock stars S2$^{\mathrm{m}}$ and S62$^{\mathrm{m}}$, generated using the parameters listed in Table~\ref{tab:dataparams}. The vertical \textit{dashed line} marks the epoch of periapsis passage, while the \textit{horizontal bars} indicate the numerically computed radial periods: $15.6\,\mathrm{yr}$ for S2$^{\mathrm{m}}$ and $9.9\,\mathrm{yr}$ for S62$^{\mathrm{m}}$.}
		\label{fig:S2S62dataplot}
	\end{figure}

	We generate two mock baseline datasets, S2$^{\rm m}$ and S62$^{\rm m}$, with orbital parameters and precision settings summarized in Table~\ref{tab:dataparams}. For both cases, we assume a Galactic Centre distance of $\Dobs = 8.2\,\mathrm{kpc}$ and a central BH mass of $\mbh = 4.3 \times 10^{6}\,M_{\odot}$, together with the PPN parameters $\gamma = \beta = 1$, a spin magnitude of $\chi = 0.5$, and spin orientation angles $\varphi_{\rm spin} = 30^{\circ}$ and $\theta_{\rm spin} = 60^{\circ}$. With these parameter choices,, the characteristic PN strength evaluates to $\epsilon \approx 2 \times 10^{-4}$ for S2$^{\rm m}$ and $\epsilon \approx 1.2 \times 10^{-3}$ for S62$^{\rm m}$.
	For the S2$^{\rm m}$ dataset, we adopt Newtonian initial conditions. In contrast, for the S62$^{\rm m}$ dataset we use 1PN-corrected initial conditions (for details, see Section~\ref{subsec:app:1PNIC}), as the periapsis is significantly closer to the BH and Newtonian initial conditions fail to accurately capture the dynamics. Observations are simulated over the time spans $T_{\rm min} = 1998.019$ to $T_{\rm max} = 2030.020$ for S2$^{\rm m}$, and $T_{\rm min} = 2000.015$ to $T_{\rm max} = 2014.005$ for S62$^{\rm m}$ (see Fig.~\ref{fig:S2S62dataplot}).

	For the orbital configurations considered (S2$^{\rm m}$ and S62$^{\rm m}$), together with the adopted observational time span and synthetic data quality tiers, the setup is not expected to yield a statistically significant detection of the BH spin. This expectation is consistent with the detectability criterion derived by \citep{waisberg2018stellar},
	\begin{equation}
		a_{\rm sma}(1-e^2)^{3/4} \lesssim 300\,r_{\rm s},
	\end{equation}
	obtained for a $4$ year observational baseline, $120$ astrometric measurements, $10\,\upmu\mathrm{as}$ astrometric precision, and a spin magnitude of $\chi= 0.9$, where $r_{\rm s}$ denotes the Schwarzschild radius.
	
\begin{figure}
		\centering
		\includegraphics[width=\columnwidth]{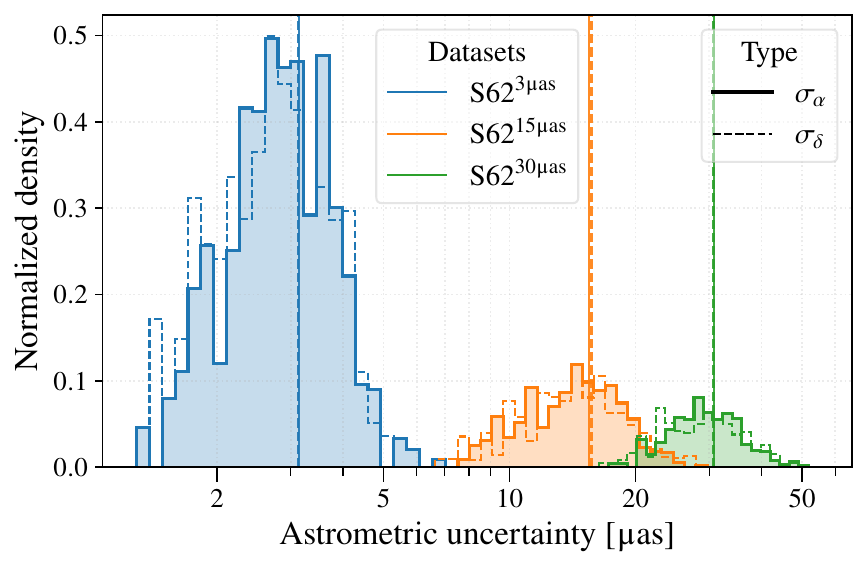}
		\caption{Distribution of astrometric uncertainties for the S62$^{3\upmu\mathrm{as}}$, S62$^{15\upmu\mathrm{as}}$ and S62$^{30\upmu\mathrm{as}}$ mock datasets.}
		\label{fig:S62highsigmas}
	\end{figure}

	In addition, we construct a set of enhanced S62-like mock datasets, denoted by S62$^{3\upmu\mathrm{as}}$, S62$^{15\upmu\mathrm{as}}$, and S62$^{30\upmu\mathrm{as}}$, with improved astrometric precisions of $\mu_{\rm ast}=3\,\upmu\mathrm{as}$, $15\,\upmu\mathrm{as}$, and $30\,\upmu\mathrm{as}$, respectively, together with slightly denser periapsis coverage, characterized by periapsis coverage fractions of $f_{\rm peri}^{\rm ast}=0.35$ and $f_{\rm peri}^{\rm RV}=0.35$. These datasets are generated assuming a high spin parameter value of $\chi = 0.9$, while the corresponding distributions of $\sigma_{\alpha}$ and $\sigma_{\delta}$ are shown in Fig.~\ref{fig:S62highsigmas}. All remaining parameters are kept identical to those of the baseline S62$^{\rm m}$ dataset, including the number of data points. This setup is intended to investigate the impact of improved measurement precision and a larger spin amplitude on parameter inference. Furthermore, to examine the impact of GL on the PPN constraints, we construct the S62$^{3\upmu\mathrm{as}}_{\rm GL}$ dataset. This dataset has the same orbital parameters, observational uncertainties, number of data points, and sampling strategy as S62$^{3\upmu\mathrm{as}}$, but the mock astrometric positions are computed including the GL contribution according to equation~\eqref{eq:GLastro}.
	
	Finally, to explore joint multi-star analyses combining an S2-like star with an enhanced S62-like star, we construct a modified $\widetilde{\mathrm{S2}}^{\rm m}$ mock dataset. This dataset is generated using the same orbital parameters and dataset-generation procedure as the baseline $\mathrm{S2}^{\rm m}$ dataset, but with the injected spin parameter increased to $\chi = 0.9$.

	
	\begin{table*}
		\centering
		\caption{Orbital parameters for S2-like and S62-like stars are adopted in this work. The BH mass and observer distance are fixed to the $\sgr$ reference values, $\mbh = 4.3 \times 10^6\, M_{\odot}$ and $\Dobs = 8.2\,\mathrm{kpc}$, respectively. 
			The orbital elements are adopted from \citealt{gillessen2009monitoring} for S2 and from \citealt{peissker2020s62} for S62.}
		\label{tab:dataparams}
		\begin{tabular}{lcccccc|cccc|cccc}
			\hline\hline
			& $a\, [\rm{arcsec}]$ & $e$ & $i \, [^{\circ}]$ & $\Omega \, [^{\circ}]$ & $\omega\,  [^{\circ}]$ & $T_{\mathrm{p}}\,[\rm{yr}]$ &  $\Delta t_{\rm p}\,[\rm{month}]$& $f_{\rm High}^{\rm ast}$  & $f_{\rm High}^{\rm RV}$ & $f_{\rm peri}^{\rm ast}$ & $f_{\rm peri}^{\rm RV}$  & $N_{\rm pos}$&$N_{\rm RV}$\\
			\hline
			S2$^{\rm m}$  & 0.125 & 0.884 & 133.82 & 227.85 & 66.13 & 2002.32 &  2.5& 0.7 & 0.7 & 0.15 & 0.15 & 93 & 46\\
			S62$^{\rm m}$ & 0.092\rlap{$^a$}   & 0.976   & 72.76   & 122.61 & 42.62 & 2003.33 & 3 & 0.7 &0.7 & 0.25 & 0.25 & 69 & 34\\
			\hline\hline
			\multicolumn{14}{@{}l@{}}{$^{a}$ \citet{peissker2020s62} report the physical semi-major axis $a_{\rm sma}$; we convert it to the angular semi-major axis using $\Dobs = 8.2\,\mathrm{kpc}$.}
		\end{tabular}
	\end{table*}
	
	\subsection{Individual contributions to astrometry and radial velocity}
	\label{sec:EffContributions}
	We separately quantify the astrometric contributions arising from the R\o mer delay, the Shapiro delay, and GL. The GL contribution is treated as a direct angular correction to the astrometric position and is described by the component shifts $\updelta\alpha_{\rm GL}$ and $\updelta\delta_{\rm GL}$, together with the total angular shift $\Delta\theta_{\rm GL}$, as defined in Section~\ref{sec:app:GL}.
	The R\o mer and Shapiro delay contributions are instead defined as time-remapping effects. In both cases, we compare unlensed astrometric positions evaluated at the same observed time, but obtained with different propagation time corrections. The precise definitions of $\Delta\theta_{\rm R}$ and $\Delta\theta_{\rm Sh}$ are given in Section~\ref{sec:app:DelayContribut}.
	
	We also quantify the radial velocity contribution associated with the Shapiro delay, denoted by $v_{\rm rad}^{\rm Sh}$. This term corresponds to the Shapiro delay contribution to the spectroscopic observable and is defined consistently with the radial velocity model in Eq.~\eqref{eq:RV}.
	
	We note that the adopted Shapiro delay and GL expressions do not depend explicitly on the spin parameter $\chi$; any spin dependence enters indirectly through the spin-dependent orbital trajectory. The resulting indirect spin dependence is negligible. Increasing the BH spin parameter from $\chi=0$ to $\chi=0.9$ changes the astrometric contribution of the Shapiro delay by only $\sim 10^{-2}\,\upmu{\rm as}$ for both orbital configurations, while the corresponding change in the GL contribution is only $\sim 10^{-3}\,\upmu{\rm as}$.
	
\begin{figure*}
		\centering
		\includegraphics[width=1\columnwidth]{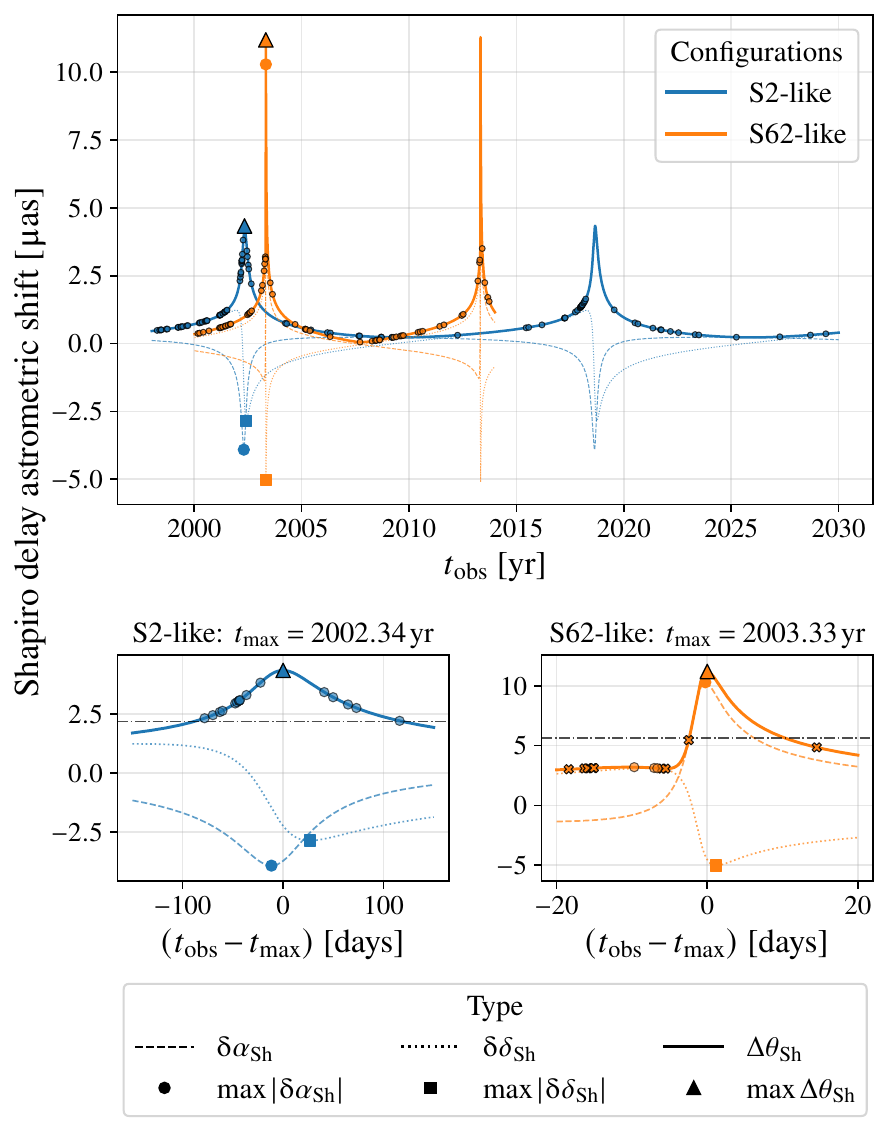}
		\includegraphics[width=1.01\columnwidth]{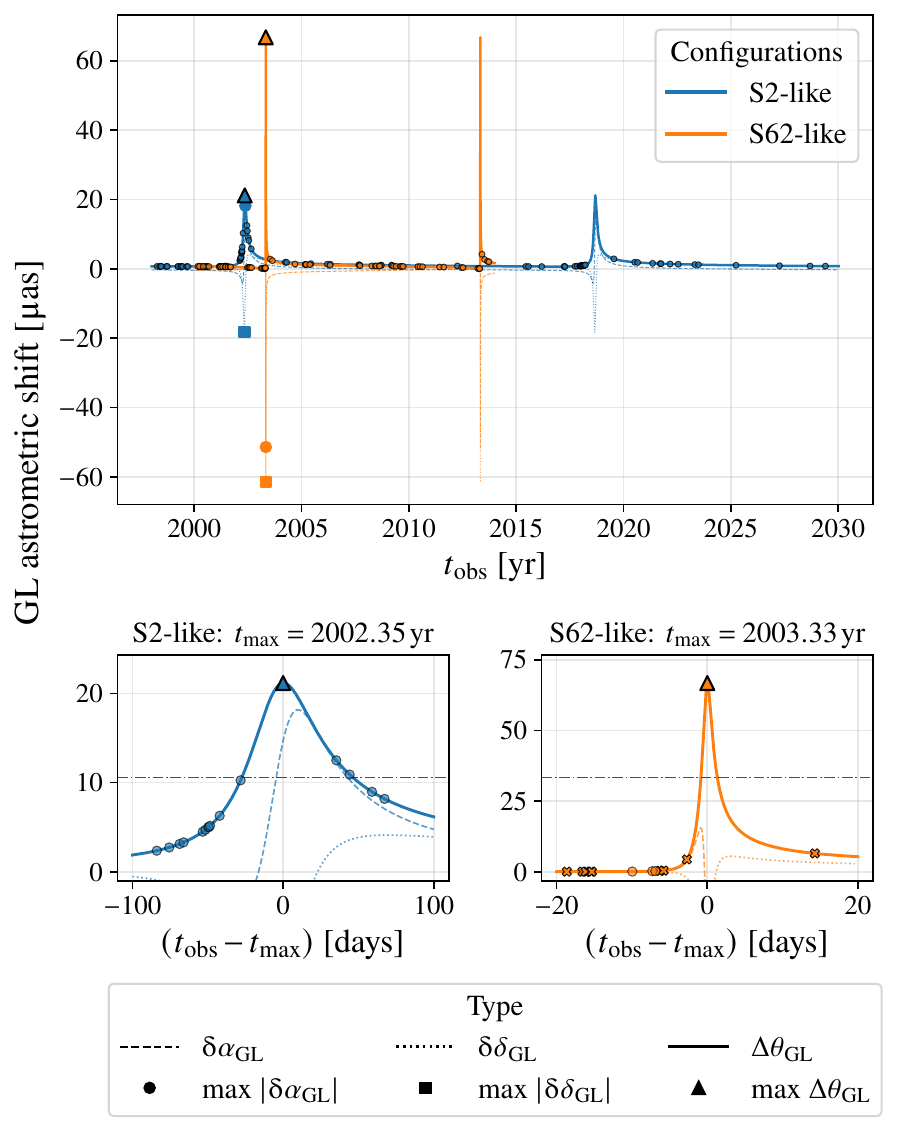}
		\caption{
			\textit{Left panel}: Astrometric contributions associated with the Shapiro time-remapping effect. \textit{Right panel}: GL correction for the S2-like and S62-like configurations. The zoomed panels in the \textit{second row} show the regions around the corresponding maxima. The \textit{dash-dotted lines} indicate $0.5$ of the maximum total contribution for each effect. \textit{Circle markers} denote the sampling of the S2$^{\rm m}$ and S62$^{\rm m}$ datasets, while \textit{cross markers} denote the enhanced S62-like datasets.}
		\label{fig:GLShapiro}
		
	\end{figure*}

\begin{figure}
		\centering
		\includegraphics[width=1\columnwidth]{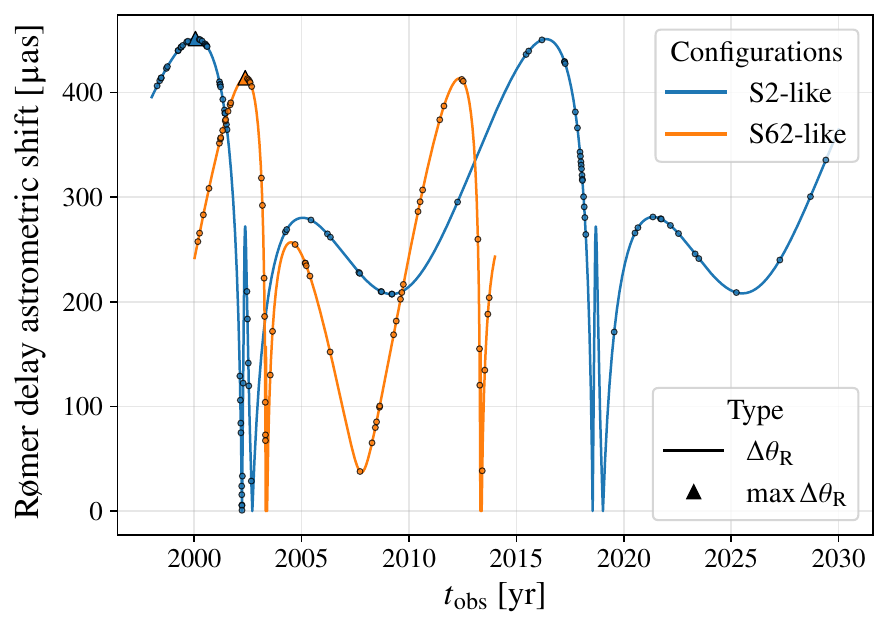}
		\caption{Astrometric contributions associated with the R\o mer time-remapping effect. \textit{Circle markers} denote the sampling of the S2$^{\rm m}$ and S62$^{\rm m}$ datasets.}
		\label{fig:Romer}
	\end{figure}

\begin{figure}
		\centering
		\includegraphics[width=1\columnwidth]{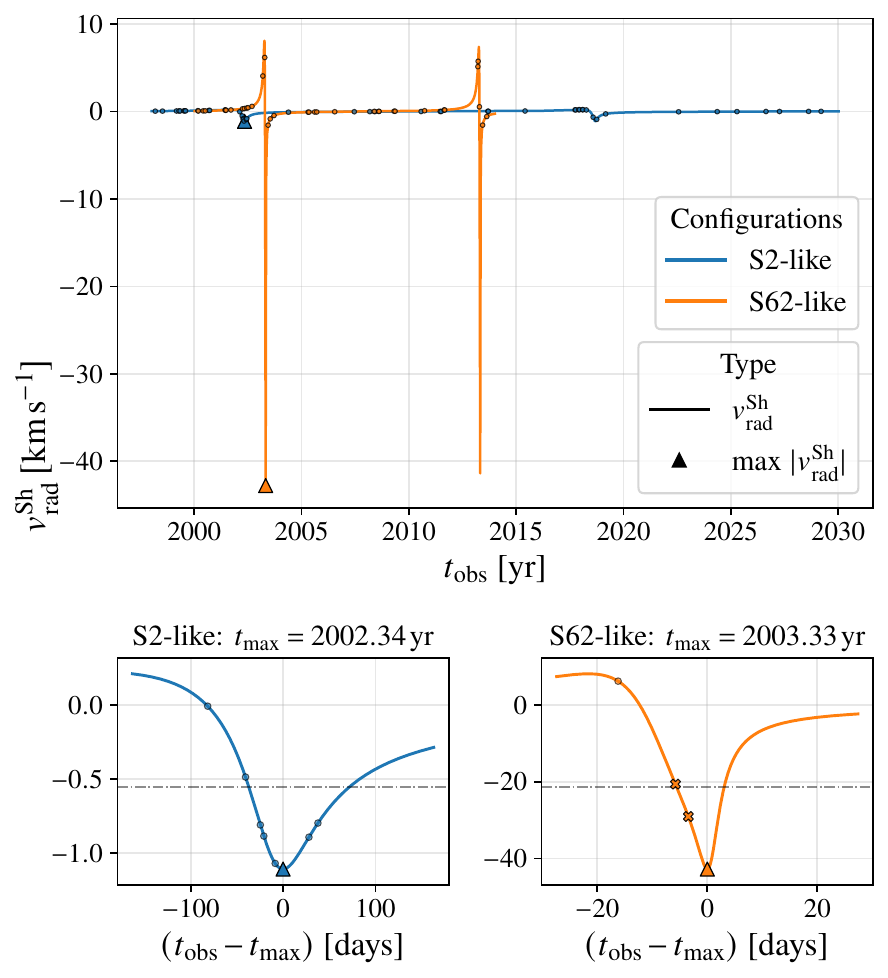}
		\caption{$v_{\rm rad}^{\rm Sh}$	for the S2-like and S62-like configurations. \textit{Circle markers} denote the sampling of the S2$^{\rm m}$ and S62$^{\rm m}$ datasets, while \textit{cross markers} denote the enhanced S62-like datasets.}
		\label{fig:RV}
	\end{figure}

	The GL and Shapiro delay contributions are presented in Fig.~\ref{fig:GLShapiro}. The R\o mer delay contribution and $v_{\rm rad}^{\rm Sh}$ are shown separately in Fig.~\ref{fig:Romer} and Fig.~\ref{fig:RV}, respectively. The sampling coverage of the different contributions for all datasets considered in this work is summarized in Table~\ref{tab:effect_sampling}.
	
	The Shapiro delay total astrometric contribution reaches approximately $4\,\upmu{\rm as}$ for the S2-like configuration and $11\,\upmu{\rm as}$ for the S62-like configuration. The corresponding GL contribution is larger, reaching approximately $21\,\upmu{\rm as}$ and $66\,\upmu{\rm as}$ for the S2-like and S62-like configurations, respectively. The R\o mer delay contribution, as a expected, is largest and reach up to $451\,\upmu\rm{as}$ for S2-like and  $414\,\upmu\rm{as}$ for S62-like configuration.
	For the baseline S2$^{\rm m}$ dataset, the largest sampled Shapiro delay contribution recovers about $88\%$ ($3.81\,\upmu\rm{as}$) of the global maximum, while the sampled GL contribution recovers about $59\%$ ($12.5\,\upmu\rm{as}$). For the baseline S62$^{\rm m}$ sampling, the recovery is poorer: about $31\%$ ($3.5\,\upmu\rm{as}$) for the Shapiro delay contribution and only about $6\%$ ($4.2\,\upmu\rm{as}$) for GL. The enhanced S62-like sampling improves the recovery of the Shapiro delay contribution to about $59\%$ ($6.7\,\upmu\rm{as}$), although the corresponding high-contribution coverage remains small, $f_{0.5}\simeq0.9\%$. For GL, the enhanced S62-like sampling recovers only about $16\%$ of the global maximum and has $f_{0.5}=0$, reflecting the very narrow temporal localization of the GL peak. 
	
	It is useful to compare the magnitudes of the individual effects shown in Figs.~\ref{fig:GLShapiro}--\ref{fig:RV} and Table~\ref{tab:effect_sampling} with the S2 analysis of \citet{grould2017general}. Their work studied the detectability of relativistic effects in mock S2 observations using a hierarchy of stellar-orbit models, ranging from Keplerian motion without R\o mer delay to a full-GR Kerr ray-tracing model. In their notation, Model~G is the most complete model and includes the relativistic stellar trajectory, R\o mer delay, Shapiro delay, GL, LT effects on the stellar orbit, and spin dependent effects on the photon trajectory. Model~F, which was used by \citet{grould2017general} for efficient spin-parameter estimation, does not perform full ray-tracing but instead incorporates an approximate weak-deflection treatment of gravitational lensing. We note that this analytical approximation is based on the formulation of \citet{sereno2006analytical} and differs from the GL model adopted in this work.
	
	The comparison of the GL contribution should be interpreted with care, because the diagnostic definitions are not identical. In \citet{grould2017general}, the GL contribution is evaluated within the full-GR ray-tracing model by comparing the lensed apparent position on the observer screen with the position obtained by projecting the same stellar spacetime event along a Euclidean straight line. This definition isolates the photon curvature effect at a fixed emission event. In the present work, instead, GL is included as a leading weak-field primary-image correction to the astrometric position. Thus, the two treatments are not mathematically identical. Nevertheless, for S2 the dominant contribution in both cases is the same physical effect: the leading weak-field mass deflection of the primary image. Spin-dependent corrections to photon propagation are expected to be negligible at this level. With this caveat in mind, the amplitudes are consistent. \citet{grould2017general} found that the GL displacement of S2 reaches a maximum of approximately $20\,\upmu{\rm as}$ near pericentre, while remaining at the $\sim2\,\upmu{\rm as}$ level over most of the orbit. The comparison of the Shapiro-delay contribution is also definition-dependent, since in \citet{grould2017general} it is inferred from the difference between the full-GR ray-traced prediction and a model in which the photon propagation does not include the Shapiro delay. Our calculation follows the same physical interpretation, namely an astrometric displacement induced by the Shapiro time remapping, but uses a leading-order PPN time-delay prescription. Therefore, the factor-of-two difference between their maximum value, approximately $8\,\upmu{\rm as}$, and our S2-like value, approximately $4\,\upmu{\rm as}$, should be interpreted as an order-of-magnitude consistency rather than a one-to-one comparison. In both calculations, the Shapiro astrometric contribution remains below the $\sim10\,\upmu{\rm as}$ level and is smaller than the GL contribution, for which our S2-like calculation gives a maximum of $21\,\upmu{\rm as}$.
	
	\begin{table}
		\centering
		\caption{
			Sampling recovery of the Shapiro delay and GL astrometric contributions, and of $v_{\rm rad}^{\rm Sh}$. Here, $t$ denotes the densely sampled reference epochs, while $t_i$ and $t_j$ denote the astrometric and radial velocity sampling epochs, respectively. The notation S62$^{+}$ denotes the enhanced S62-like datasets, namely S62$^{30\upmu{\rm as}}$, S62$^{15\upmu{\rm as}}$ and S62$^{3\upmu{\rm as}}$. The quantity $f_{0.5}$ is the fraction of sampled epochs satisfying $\Delta\theta(t_i)>0.5\mathop{\max}\limits_{\tiny t}\left( \Delta\theta\right)$. All quantities are evaluated over the full time series; therefore, when two comparable peaks are present, sampled epochs near either peak contribute to the reported values.
		}
		\label{tab:effect_sampling}
		\begin{tabular}{llccc}
			\hline\hline
			Effect
			& Dataset
			&  $\mathop{\max}\limits_{\tiny t}\left( \Delta\theta\right)\,[\upmu\rm{as}]$
			& $\mathop{\max}\limits_{\tiny t_i}\left( \Delta\theta\right)$
			& $f_{0.5}\,[\%]$
			\\
			\hline
			
			\multirow{3}{*}{Shapiro}
			& S2$^{\rm m}$      & 4.34  & 3.81  & 17.0 \\
			& S62$^{\rm m}$     & 11.29 & 3.50  & 0.0  \\
			& S62$^{\rm +}$ & - & 6.67  & 0.9  \\
			
			\cline{1-5}
			
			\multirow{3}{*}{GL}
			& S2$^{\rm m}$      & 21.16 & 12.50  & 2.2 \\
			& S62$^{\rm m}$     & 66.75 & 4.18   & 0.0 \\
			& S62$^{\rm +}$ & - & 10.46  & 0.0 \\
			
			\hline

			& 
			& $\mathop{\max}\limits_{\tiny t}\left|v_{\rm rad}^{\rm Sh}\right|\,[{\rm km\,s^{-1}}]$
			& $\mathop{\max}\limits_{\tiny t_j}\left|v_{\rm rad}^{\rm Sh}\right|$ 
			&
			\\
			\hline
			
			\multirow{3}{*}{RV}
			& S2$^{\rm m}$      & 1.10 & 1.06  & 15.2 \\
			& S62$^{\rm m}$     & 42.78& 6.20  & 0.0 \\
			& S62$^{\rm +}$ & - & 36.56  & 0.1 \\
			
			\hline
		\end{tabular}
	\end{table}
	
	\section{Parameter Inference}\label{seq:Inference}
	\subsection{Parameter Space and Observables}
	In our physical model, the parameter set is given by $\Theta = \{\mbh,\Dobs,a,e,\iota,\Omega,\omega,T_{\rm p},\chi,\theta_{\rm spin},\varphi_{\rm spin},\gamma,\beta\}$. These parameters can be grouped as follows:
	\begin{itemize}
		\item ${\mbh, \chi, \gamma, \beta}$ define the underlying spacetime and the gravitational theory,
		\item ${\theta_{\rm spin}, \varphi_{\rm spin}, \Dobs}$ specify the orientation of the system and the overall observational scaling,
		\item ${a, e, \iota, \Omega, \omega, T_{\rm p}}$ parameterize the orbital initial conditions within this spacetime.
	\end{itemize}
	
	Here, $\theta_{\rm spin}$ and $\varphi_{\rm spin}$ define the orientation of the BH spin axis relative to the observer. While they do not alter the intrinsic structure of the spacetime (which is fully specified by $\mbh$ and $\chi$), they determine how the geometry is probed by the orbit and projected onto observables
	
	The observables dataset consist of astrometric and radial velocity mock measurements:
	\begin{equation}
		\mathcal{D} = \left\{(\Delta \alpha_{\rm obs}(t_i), \Delta \delta_{\rm obs}(t_i)), \;
		v_{\rm rad,obs}(t_j)\right\}.
	\end{equation}
	
	\subsection{Parameter degeneracies}\label{subsec:DegCorr}
	Our parameter space is $13$-dimensional and exhibits several important degeneracies and correlations. At the Keplerian level, purely astrometric observations constrain only the combination $\mbh \propto \Dobs^3$, because astrometry measures $a=a_{\rm sma}/\Dobs$ while the orbital dynamics determine $a_{\rm sma}^3/\mbh$ \citep{Gillessen2017}. Radial velocity measurements break this degeneracy by providing an absolute physical velocity scale. Residual correlations also persist between the inclination $\iota$ and the angular semi-major axis $a$.
	
	At the 1PN level, a degeneracy arises between the PPN parameters $\gamma$ and $\beta$ with respect to the Schwarzschild pericentre precession equation~\eqref{eq:SchPrec}, as they enter the equation strictly through the linear combination $\Upsilon$ \citep{de2025future}. Because pericentre precession is typically the dominant and most readily detectable relativistic effect in stellar orbits, observational data primarily constrain this $\Upsilon$ combination rather than the individual parameters. In principle, this degeneracy can be partially broken by relativistic observables that depend differently on the PPN parameters, such as the Shapiro time delay equation~\eqref{eq:Delays} and the relativistic radial velocity corrections equation~\eqref{eq:RV} and also GL equation~\eqref{eq:GLgamma}. However, these secondary effects are often strongly suppressed depending on the orbital configuration and measurement precision, so the data remain predominantly sensitive to the $\Upsilon$ combination in practice.
	Furthermore, a correlation exists between $\gamma$ and $\mbh$, since the 1PN dynamics are governed by the effective mass $(1+\gamma)\mbh$. However, because $\mbh$ dictates the underlying orbital mechanics already at the Keplerian level, its value is strongly anchored by the combined astrometric and radial velocity data. This Keplerian dominance inherently restricts the $\gamma$--$\mbh$ correlation, as any significant variation in $\gamma$ to absorb mass uncertainties would disrupt the primary Keplerian fit.
	
	At the 1.5PN order, additional correlations emerge involving the BH spin. The LT precession, given in equation~\eqref{eq:LTomegaPrec}, scales as $\chi(1+\gamma)\epsilon^{3/2}$, introducing couplings between the spin magnitude $\chi$, the PPN parameter $\gamma$, and the orbital parameters through the spin-orbit projection terms and the PN strength parameter $\epsilon$. Since the LT contribution enters at higher PN order than the dominant Schwarzschild-like precession, its observational impact is generally weaker for $\epsilon\ll1$. Consequently, correlations between $\chi$ and the orbital elements, such as the semi-major axis $a_{\rm sma}$ and eccentricity $e$, are expected to remain subdominant.
	
	A particularly important effect is the emergence of a degeneracy between the effective Schwarzschild precession parameter $\Upsilon$ and the spin parameter $\chi$. Although the spin-orbit contribution formally depends on the combination $\chi(1+\gamma)$, the dominant observational degeneracy is not directly between $\gamma$ and $\chi$, but rather between $\Upsilon$ and $\chi$. This behavior follows from the hierarchy of PN contributions to the total apsidal precession: $\Delta\omega_{\rm 1PN} \propto \epsilon\,\Upsilon$ and $\Delta\omega_{\rm 1.5PN} \propto \epsilon^{3/2}\chi(1+\gamma)$.
	Because the Schwarzschild-like precession dominates the relativistic signal, variations in the weaker LT contribution can be partially compensated by shifts in $\Upsilon$.
	
	This degeneracy becomes particularly severe when the astrometric time span does not cover multiple periapsis passages. In such cases, the secular out-of-plane signatures of LT frame dragging, including nodal and inclination precession, do not accumulate sufficiently to be resolved above the astrometric noise. The fit then becomes dominated primarily by the total in-plane apsidal shift generated during periapsis passage, rendering the individual 1PN and 1.5PN contributions observationally difficult to disentangle.
	
	The dominant $\Upsilon$--$\chi$ degeneracy direction can be obtained by requiring the total relativistic apsidal precession to remain approximately constant, $\delta(\Delta\omega_{\rm tot}) \approx 0$. To leading order, this condition yields an approximately linear degeneracy relation,
	\begin{equation}\label{eq:Upschidirection}
		\Upsilon(\chi) \simeq \mathcal{P}\chi + \mathcal{C},
	\end{equation}
	where the slope $\mathcal{P}$ depends on the spin-orbit geometry and the PPN parameter $\gamma$. The intercept $\mathcal{C}$ depends on the fiducial point about which the degeneracy relation is linearized and therefore inherits contributions from the injected values of both $\Upsilon$ and $\chi$. For configurations centered near the GR solution, $\mathcal{C}$ is expected to remain close to unity.
	
	Similarly, the correlations between the spin magnitude $\chi$ and its orientation angles can be obtained by requiring the 1.5PN LT contribution to the precession to remain constant, such that $\delta(\Delta\omega_{\rm 1.5PN}) \approx 0$. By isolating the variations of the polar and azimuthal angles independently, we can find two primary degeneracy boundaries. For a fixed azimuthal angle, the degeneracy in the $(\chi,\,\theta_{\rm spin})$ plane follows a phase shifted secant curve:
	\begin{equation}\label{eq:thetachidirection}
		\chi(\theta_{\rm spin}) \simeq \mathcal{K}_{\theta} \sec(\theta_{\rm spin} - \alpha).
	\end{equation}
	Here, the phase shift $\alpha$ is determined entirely by the projection of $\varphi_{\rm spin}$ onto the orbital geometry, while the normalization coefficient $\mathcal{K}_{\theta}$ combines the overall amplitude of the observed 1.5PN signal with the corresponding geometric projection factors. Because the secant function diverges rapidly as $\theta_{\rm spin}$ deviates from $\alpha$, the physical bound $\chi \le 1$ truncates the posterior, resulting in constrained polar angle estimates for high spin configurations. Conversely, for a fixed polar angle, the degeneracy in the $(\chi,\,\varphi_{\rm spin})$ plane traces an inversely modulated sine wave:
	\begin{equation}\label{eq:phichidirection}
		\chi(\varphi_{\rm spin}) \simeq \frac{\mathcal{K}_{\varphi}}{\mathcal{A} + \mathcal{B} \sin(\Omega - \varphi_{\rm spin})},
	\end{equation}
	where the baseline shift $\mathcal{A}$ and the modulation amplitude $\mathcal{B}$ are geometric coefficients completely determined by the value of $\theta_{\rm spin}$ while $\mathcal{K}_{\varphi}$ acts as integration constants that represent the absolute physical magnitude of the 1.5PN signal required by the observational data.
	
	Furthermore, the spin orientation angles themselves can be mutually correlated~\citep{anglesdeg,grould2017general}; however, in our datasets the spin angles do not exhibit strong correlations, as shown in Fig.~\ref{fig:precrv}.
	
	\subsection{Likelihood and priors}\label{sec:LikelihoodPriors}
	\subsubsection{The joint likelihood}
	To perform the parameter inference, we construct a joint log-likelihood function, $\ln \mathcal{L}(\Theta\given\mathcal{D}) = \ln \mathcal{L}_{\mathrm{pos}}(\Theta\given\mathcal{D}) + \ln \mathcal{L}_{\mathrm{RV}}(\Theta\given\mathcal{D})$, which sums the independent contributions from the astrometric positions and the radial velocities:
	\begin{align}\label{eq:MCMCLL}
		\ln \mathcal{L}_{\mathrm{pos}}(\Theta\given\mathcal{D}) &= -\frac{1}{2} \sum_{i=1}^{N_{\mathrm{pos}}} \left[ (\myvec{d}_i - \hat{\myvec{d}}_i)^{\mathrm{T}} \mymatrix{C}_i^{-1}  (\myvec{d}_i - \hat{\myvec{d}}_i) + \ln(|2\pi\mymatrix{C}_i|) \right],\\
		\ln \mathcal{L}_{\mathrm{RV}}(\Theta\given\mathcal{D}) &= -\frac{1}{2} \sum_{j=1}^{N_{\mathrm{RV}}} \left[ \frac{(v_{\mathrm{rad, obs}, j} - \hat{v}_{\mathrm{rad}, j})^2}{\sigma_{\mathrm{RV}, j}^2} + \ln(2\pi \sigma_{\mathrm{RV}, j}^2) \right],
	\end{align}
	where $\myvec{d}_i := (\Delta \alpha_{\mathrm{obs}}(t_i), \Delta \delta_{\mathrm{obs}}(t_i))^{\mathrm{T}}$ is the observed position vector at time $t_i$, $\hat{\myvec{d}}_i$ is the model-predicted position vector derived from the parameters $\Theta$, $\mymatrix{C}_i$ is the measurement covariance matrix defined in equation~\eqref{eq:adeltacov}, and $\hat{v}_{\mathrm{rad}, j}$ is the model-predicted radial velocity.
	
	\subsubsection{Priors}
	Our primary MCMC inference explores the physical parameter space directly. The global prior probability $p(\Theta)$ is constructed as the product of independent uniform priors for each physical parameter. The adopted prior ranges are listed in the last row of Table~\ref{tab:MAPparametrization}. For the PPN framework, to mitigate the correlations discussed in Section~\ref{subsec:DegCorr}, we replace the standard $(\gamma, \beta)$ parameters with the $(\Upsilon,\gamma)$ parameterization, adopting uniform priors $\mathcal{U}(0.9, 1.1)$ for both. The dimensionless spin magnitude is similarly bounded by $\chi \sim \mathcal{U}(0,1)$, with uniform priors over the sphere for the spin orientation angles.
	
	\subsubsection{MAP likelihood modifications and parametrization}\label{subsubsec:MAPMod}
	While the full joint likelihood and physical priors described above are used for the final MCMC inference, we apply specific simplifications to this mathematical framework to construct a well-conditioned target function for our initial MAP optimization stage. During the MAP estimation, we minimize a globally normalized (tempered) negative log-posterior:
	
	\begin{equation}\label{eq:LMAP}
		\mathcal{L}_{\rm MAP}(\Theta,\mathcal{D})
		=
		-\frac{1}{N_{\rm tot}}
		\left[
		\ln \widetilde{\mathcal{L}}
		(\Theta\given\mathcal{D})
		+
		\ln p(\Theta)
		\right],
	\end{equation}
	where $\widetilde{\mathcal{L}}$ denotes the simplified likelihood adopted during the preliminary MAP optimization, and
	$N_{\rm tot}=2N_{\rm pos}+N_{\rm RV}$ is the total number of observables. This rescaling does not alter the location of the MAP solution but improves the numerical conditioning of the optimization. Because gradient-based optimization is highly sensitive to hard prior boundaries, the MAP estimation is carried out in an unconstrained parameter space by introducing transformed (\textit{raw}) variables. This reparameterization, summarized in Table~\ref{tab:MAPparametrization}, guarantees that all physical constraints are automatically satisfied while preserving smooth gradients. Finally, to isolate the Keplerian and orientation parameters during this warmup phase, we fix the PPN parameters to their GR values ($\gamma = \beta = 1$) and the spin parameter to $\chi = 0$. We also adopt a simplified positional likelihood during this stage that assumes independent Gaussian uncertainties for $\Delta\alpha_{\rm obs}$ and $\Delta\delta_{\rm obs}$ (for details, see Section~\ref{sec:app:MAPDet}).
	
	\begin{table*}
		\centering
		\caption{Model parameterisation for MAP estimation. All parameters are optimised in an unconstrained space $\tilde{x} \in \mathbb{R}$ and mapped to physical parameters via deterministic transformations. The sigmoid function is defined as $\sigma(\tilde{x}) = (1 + e^{-\tilde{x}})^{-1}$. The time of pericentre is parameterised relative to a reference epoch as $\Tp = T_{\rm ref} + \Delta \Tp$, where $T_{\rm ref} = 2002.0$ for S2$^{\rm m}$ and $T_{\rm ref} = 2003.0$ for S62$^{\rm m}$. The BH mass $\mbh$ is expressed in units of $10^6\,M_\odot$, distances $\Dobs$ in kpc, angular semi-major axis $a$ in arcseconds, and $\Delta\Tp $ in years. Priors are defined in the physical parameter space and provided in last column. Here, $\mathrm{LogNormal}(\mu,\sigma)$ denotes a log-normal distribution parameterized by mean and standard deviation in logarithmic space.}
		\label{tab:MAPparametrization}
		\begin{tabular}{l|lll}
			\hline
			Physical parameter & Raw variable & Transformation & Prior \\
			\hline
			
			$\mbh\,[10^6\,M_\odot]$ 
			& $\tilde{\mu}$ 
			& $ \mbh = e^{\tilde{\mu}}\,\Dobs^3$ 
			& $\mathrm{LogNormal}\big(\ln(4),\,0.2\big)$ \\
			
			$\Dobs\,[\rm{kpc}]$ 
			& $\tilde{D}$ 
			& $\Dobs = e^{\tilde{D}}$ 
			& $\mathrm{LogNormal}\big(\ln(8),\,0.2\big)$ \\
			
			$a\,[\rm{arcsec}]$ 
			& $\tilde{a}$ 
			& $a = e^{\tilde{a}}\Dobs^{-1}$ 
			& $\mathrm{LogNormal}\big(\ln(0.1),\,0.5\big)$ \\
			
			$e$ 
			& $\tilde{e}$ 
			& $e = 0.99\,\sigma(\tilde{e})$ 
			& $ \mathcal{U}(0,\,0.99)$ \\
			
			$i\, [\rm{rad}]$ 
			& $\tilde{i}$ 
			& $i = \arccos\left(2\sigma(\tilde{i})-1\right)$ 
			& $\mathcal{U}(0,\,\pi)$ \\
			
			$\Omega\, [\rm{rad}]$ 
			& $\tilde{\Omega}$ 
			& $\Omega = \mathrm{mod}(\tilde{\Omega},\,2\pi)$ 
			& $\mathcal{U}(0,\,2\pi)$ \\
			
			$\omega\, [\rm{rad}]$ 
			& $\tilde{\omega}$ 
			& $\omega = \mathrm{mod}(\tilde{\omega},\,2\pi)$ 
			& $\mathcal{U}(0,\,2\pi)$ \\
			
			$\Delta \Tp\,[\mathrm{yr}]$ 
			& $\tilde{\Delta \Tp}$ 
			& $\Delta \Tp = T_{\min} + (T_{\max}-T_{\min})\,\sigma(\tilde{\Delta \Tp})$ 
			& $\mathcal{U}(T_{\min}-T_{\rm ref},\,T_{\max}-T_{\rm ref})$  \\
			
			\hline
		\end{tabular}
	\end{table*}
	
	\subsection{Two stage inference}
	Our parameter inference framework is implemented in two sequential stages designed to address the high dimensionality and complex degeneracies of the orbital parameter space. First, we perform a MAP optimization using the baseline datasets S2$^{\rm m}$ and S62$^{\rm m}$ to identify a high-posterior-probability region of the parameter space. Second, the resulting optimized parameter set is used to initialize the full MCMC sampling procedure for all datasets.
	We emphasize that the two-stage procedure used here does not correspond to a different statistical model from a conventional one-stage Bayesian inference. The posterior sampled in the second stage is the full posterior defined by the likelihood and priors in Section~\ref{sec:LikelihoodPriors}. The MAP stage is used only to provide a numerically efficient initialization for the parameters, whose broad prior ranges would otherwise make the NUTS warm-up less efficient. In contrast, a one-stage inference would initialize the NUTS chains directly from the prior and allow the sampler itself to locate the typical set. Both approaches therefore target the same posterior distribution; the difference is computational rather than statistical. To avoid imposing the MAP solution on the relativistic sector, the spin magnitude, spin orientation angles, and PPN parameters are initialized from their global priors at the beginning of the MCMC warm-up.
	
	\subsubsection{MAP optimization warmup}\label{subsec:MAPWarmup}
	With the modified target function and unconstrained parameter space established in Section~\ref{subsubsec:MAPMod}, our initial optimization stage aims to rapidly lock onto the system's Keplerian and orientation parameters. We minimize the tempered negative log-posterior, $\mathcal{L}_{\rm MAP}(\Theta,\mathcal{D})$, using the Adam optimizer \citep{kingma2014adam}.
	
	Once the optimizer converges to a stable minimum, the resulting unconstrained \textit{raw} parameters are deterministically mapped back into the physical parameter space using the transformations listed in Table~\ref{tab:MAPparametrization}. These best-fit values are then used to initialize the subsequent MCMC sampling phase. 
	
	To reduce sensitivity to the initial conditions and mitigate the possibility of convergence to a suboptimal local minimum, the optimization is repeated from multiple initial parameter sets randomly drawn from the prior distributions. Among the resulting solutions, the one with the highest posterior probability (equivalently, the lowest negative log-posterior) is selected as the initialization for the subsequent MCMC sampling.

	\subsubsection{MCMC posterior sampling}\label{subsec:MCMCSampling}
	While MAP estimation provides a single best-fit point in parameter space corresponding to the peak of the posterior distribution, and is therefore useful for obtaining a representative solution, it has several important limitations. In particular, MAP estimation does not capture the full structure of the posterior, provides no direct information about parameter uncertainties or correlations, and may be sensitive to local extrema, especially in high-dimensional or multimodal parameter spaces. As a result, there is no guarantee that the MAP solution adequately represents the full range of plausible models consistent with the data. To fully characterize the posterior distribution and quantify parameter uncertainties and correlations, we therefore proceed to MCMC sampling.
	
	We implement our MCMC framework using the {\texttt{NumPyro}\footnote{\url{https://github.com/pyro-ppl/numpyro}} library \citep{phan2019composable,bingham2019pyro}, employing the No-U-Turn Sampler (NUTS) \citep{hoffman2014no}. As an advanced variant of Hamiltonian Monte Carlo (HMC), NUTS utilizes gradient information from the posterior landscape to automatically adapt its trajectory length, enabling efficient exploration of high-dimensional parameter spaces without manual tuning of the integration path. In this stage, we use the log-likelihood defined in equation~\eqref{eq:MCMCLL}. Since the PPN and spin parameters are fixed during the MAP warmup stage, their initial values for the MCMC sampling are drawn directly from the corresponding uniform priors.
		
		For every evaluation run, we deploy $10$ independent chains. Each chain executes $2000$ warmup steps to optimize the mass matrix and step size, followed by $5000$ sampling steps, yielding a total pool of $50000$ posterior samples per run. 
		
		We assess chain convergence using the Gelman--Rubin diagnostic, $\hat{R}$ \citep{gelman1992inference}. For the vast majority of physical parameters, we achieve tight convergence with $\hat{R} \lesssim 1.01$. A slight elevation up to $\hat{R} \sim 1.05$ is confined entirely to the highly degenerate $(\Upsilon, \chi)$ sector; this behavior is physically expected due to the elongated, banana-shaped geometry of the posterior covariance along this manifold, which naturally yields longer structural autocorrelation times. To guarantee that these regions are still rigorously mapped, we monitor both the bulk and tail effective sample sizes, verifying that $\mathrm{ESS}_{\mathrm{bulk}}, \mathrm{ESS}_{\mathrm{tail}} \gtrsim 1000$ across all parameters. Crucially, the total absence of post-warmup divergent transitions confirms that the sampler has explored the intricate geometric structures of our posterior.
		
		\subsection{Inference runs and experimental configurations}\label{subsec:Configurations}
		
		We consider the following inference experiments:
		
		\begin{enumerate}
			\item {Baseline Analysis:} We explore the full parameter space using the baseline S2$^{\mathrm{m}}$ and S62$^{\mathrm{m}}$ datasets to establish the reference constraints.
			\item {Enhanced Astrometry Scenarios:} To evaluate the impact of improved astrometric precision, we perform independent inference runs for three upgraded astrometric uncertainty tiers: S62$^{3\upmu\mathrm{as}}$, S62$^{15\upmu\mathrm{as}}$ and S62$^{30\upmu\mathrm{as}}$. For these high-precision runs, the background spacetime parameters (mass $\mbh$ and distance $\Dobs$) are fixed to their injected values. Each tier is analysed under two sub-scenarios in order to separate spin detectability from PPN-spin degeneracy.
			\begin{itemize}
				\item The PPN parameters are frozen to their injected GR values ($\gamma=\beta=1$), forcing the sampler to fit remaining residuals purely with spin.
				The fixed-PPN case is not used as the final physical model; rather, it serves as a control experiment defining the optimistic GR baseline for spin recovery. If the spin cannot be recovered even when the PPN sector is fixed, then the data do not contain sufficient LT information.
				\item The PPN parameters are treated as completely free parameters, allowing us to quantify how strongly variations in $\Upsilon$ and $\gamma$ can mimic or absorb spin induced effects.
			\end{itemize}
			\item {GL scenario:} To quantify the impact of GL on the PPN inference, we compare the free-PPN posterior constraints obtained from the S62$^{3\upmu\mathrm{as}}$ and S62$^{3\upmu\mathrm{as}}_{\rm GL}$ datasets. In both cases, the sampling is performed in terms of $(\Upsilon,\gamma)$, while $\beta$ is computed deterministically.
			\item {Fixed-Spin Scans:} Using the highest-precision S62$^{3\upmu\mathrm{as}}$ dataset, we perform a dedicated grid analysis. We fix the dimensionless spin magnitude across a sequence of discrete values, $\chi_{\rm fixed} \in [0, 1]$ in steps of 0.1. At each step, we infer the remaining orbital and PPN parameters to directly map the structural shifts in parameter correlations as a function of spin magnitude.
			\item {Multi-star Inference:} Using the S62$^{3\upmu\mathrm{as}}$ dataset together with the modified $\widetilde{\mathrm{S2}}^{\rm m}$ dataset to examine the reduction of the $\Upsilon$--$\chi$ degeneracy.
		\end{enumerate}
		
		For all enhanced and isolated star configurations, the MCMC chains utilize the coordinate positions from the joint baseline MAP optimization as their geometric initialization. However, to prevent optimization bias from artificially restricting the new physics sectors, the spin magnitude $\chi$, its orientation angles, and the PPN metrics are always drawn randomly from their global prior distributions at the start of the MCMC warmup phase.

		\section{Results and discussion}\label{seq:results}
		The posterior distributions inferred from the baseline S2$^{\rm m}$ and S62$^{\rm m}$ datasets are shown in Fig.~\ref{fig:S2corner} and Fig.~\ref{fig:S62corner}, respectively. Both datasets yield recoveries of the orbital parameters and the BH mass-distance combination, with the injected values falling well within the $1\sigma$ credible intervals. However, the S2$^{\rm m}$ dataset exhibits a correlation between the angular semi-major axis $a$ and eccentricity $e$. This correlation occurs because S2$^{\rm m}$ has a longer orbital period; although its pericenter is strongly constrained, the incomplete orbital coverage leaves $a$ and $e$ partially entangled. 
		
		For both the S2$^{\rm m}$ and S62$^{\rm m}$ datasets, the PPN parameter constraints are effectively obtained on the linear combination $\Upsilon$ rather than independently resolving the parameter pair $(\Upsilon, \gamma)$. For the S2$^{\rm m}$ dataset, the inferred value is $\Upsilon = 1.012 \pm 0.035$. For the S62$^{\rm m}$ dataset, the posterior mean exhibits a deviation from the injected value comparable to that of S2$^{\rm m}$, while the $1\sigma$ credible interval is slightly tighter, yielding $\Upsilon = 0.987 \pm 0.020$. As anticipated from the previously discussed parameter degeneracies and the dominance of the Keplerian contribution, the posterior distribution for $\gamma$ remains largely prior dominated, with only weak constraints from the data.
		The inability to independently isolate $\gamma$ becomes physically apparent when comparing the magnitude of its secondary relativistic effects to the assumed observational precision. Because the GL contribution is not included in the S2$^{\rm m}$ and S62$^{\rm m}$ datasets, independent sensitivity to the parameter $\gamma$ is provided by the Shapiro delay and $v_{\rm rad}^{\rm Sh}$. As shown in Section~\ref{sec:EffContributions} These effects require high-precision astrometry for S2$^{\rm m}$ and denser periapsis coverage, particularly for S62$^{\rm m}$. Lacking an exceptionally dense observational cadence exactly during this brief window, the peak signal is missed. Any remaining subtle off-peak signatures are subsequently absorbed by the strongly anchored Keplerian parameters, ensuring the observational data remain predominantly sensitive to the secular precession governed by $\Upsilon$.
		
		The results obtained for the enhanced S62-like datasets illustrate the interplay between spin effects and the PPN sector. First, we note that across all analyzed configurations, regardless of whether the PPN parameters are fixed or allowed to vary simultaneously, the orbital elements (e.g., $a, e, i, \Omega, \omega$) remain exceptionally well constrained Table~\ref{tab:orbitalsigma}. This recovery arises because the global orbital geometry is governed primarily by the dominant Keplerian dynamics and leading order kinematics, which are tightly anchored by the high-precision astrometric data and remain largely decoupled from higher order relativistic perturbations.
		In contrast to the baseline datasets, where the $\gamma$ posterior is effectively uniform, the enhanced S62-like configurations reveal a distinct peak in the distribution, especially for S62$^{3\upmu\mathrm{as}}$. However, $\gamma$ remains weakly constrained, exhibiting a significantly wider $1\sigma$ credible interval compared to that of $\Upsilon$.  (see Fig.~\ref{fig:chimodels}, right panel, for the fixed-spin case).
		
		We also tested whether the inclusion of the GL astrometric contribution improves the independent constraint on $\gamma$. To this end, we repeated the free-PPN inference using the S62$^{3\upmu\mathrm{as}}_{\rm GL}$ dataset while fixing the BH spin parameter to its injected value, $\chi=0.9$, thereby isolating the impact of the GL signal from the spin--PPN degeneracy. The resulting posterior was then compared with the corresponding fixed-spin inference for the S62$^{3\upmu\mathrm{as}}$ dataset (i.e.\ the $\chi_{\rm fixed}=0.9$ case shown in Fig.~\ref{fig:chimodels}).
		
		The resulting constraint on $\Upsilon$ remains essentially unchanged, with $\sigma_{\Upsilon}\approx10^{-4}$ in both cases. Likewise, the constraint on $\gamma$ changes only marginally, from $\gamma\simeq1.037\pm0.03$ to $\gamma\simeq1.028\pm0.03$ when the GL contribution is included. The corresponding posterior distributions are shown in Fig.~\ref{fig:Gammagamma}. Thus, even after removing the $\Upsilon$--$\chi$ degeneracy, the inclusion of GL does not significantly improve the independent constraint on $\gamma$ for the adopted observing cadence and noise level. Although the GL signal depends directly on the light-deflection factor $(1+\gamma)$, its constraining power remains limited because only a small fraction of the strongest lensing signal is sampled by the observations, while the overall GL signal is weak relative to the adopted astrometric precision.
		
		When the PPN parameters are fixed to their injected GR values, the spin parameter can be partially constrained, although degeneracies persist between the spin magnitude and the spin orientation angles. As expected, these spin constraints improve with increasing astrometric precision (first row of Fig.~\ref{fig:S62+exp}). For the S62$^{3\upmu\mathrm{as}}$ dataset, the injected spin magnitude, $\chi = 0.9$, is recovered with a credible interval of $\sigma_{\chi}\approx 0.008$. For the S62$^{15\upmu\mathrm{as}}$ and S62$^{30\upmu\mathrm{as}}$ datasets, the uncertainty increases to approximately $\sigma_{\chi}\approx 0.04$ and $\sigma_{\chi}\approx 0.06$, respectively. The spin orientation angles $(\theta_{\rm spin},\varphi_{\rm spin})$ exhibit broader credible intervals, especially for S62$^{15\upmu\mathrm{as}}$ and S62$^{30\upmu\mathrm{as}}$, reflecting the weaker sensitivity of the observables to the exact spin direction and the degeneracies between the spin angles and $\chi$ (shown in Fig.~\ref{fig:chiangles} for the S62$^{15\upmu\mathrm{as}}$ dataset).
		Degeneracy direction coefficients appearing in equation~\eqref{eq:phichidirection} and equation~\eqref{eq:thetachidirection} can be estimated by substituting the injected orbital parameters, BH mass, and distance, together with the adopted isotropic priors on the spin orientation angles. This yields the ranges $\mathcal{A} \in [-0.889,\,0.889], \, \mathcal{B} \in [0,\,1.818], \, \alpha \in [-63.95^\circ,\,63.95^\circ]$. Furthermore, by requiring the 1.5PN contribution to remain bounded by the maximum precession producible by an extremal BH ($\chi=1$), limits can be placed on the corresponding amplitude coefficients. The azimuthal amplitude is bounded within $\mathcal{K}_{\varphi} \in [-2.024,\,2.024]$, while the effective polar amplitude spans the unconstrained mathematical range $\mathcal{K}_{\theta} \in [-2.277,\,2.277]$.
		
		When the PPN parameters are allowed to vary simultaneously with the spin parameter (second row of Fig.~\ref{fig:S62+exp}), the posterior becomes dominated by the $\Upsilon$--$\chi$ degeneracy discussed in Section~\ref{subsec:DegCorr}. The left panel of Fig.~\ref{fig:cornerPPNchi} shows the resulting elongated posterior structure, while the orbital elements themselves remain well constrained (Table~\ref{tab:orbitalsigma}). Because the LT contribution is subdominant, small fluctuations in the mock astrometry permit shifts in $\Upsilon$ that can be compensated by changes in $\chi$, driving the posterior toward the physical prior boundary at $\chi=1$. Since the mock datasets are generated using the same physical model employed in the inference, this edge accumulation is not caused by unmodeled physics, but instead reflects the degeneracy between $\Upsilon$ and $\chi$.
		
		For the S62$^{3\upmu{\rm as}}$ dataset, the fixed spin experiments shown in Fig.~\ref{fig:chimodels} illustrate the underlying $\Upsilon$--$\chi$ degeneracy explicitly. Increasing the imposed spin magnitude from $\chi=0$ to $\chi=1$ shifts the inferred value of $\Upsilon$ monotonically from $\Upsilon\simeq0.961$ to $\Upsilon\simeq1.005$, while the corresponding variations in $\gamma$ remain comparatively weak. This behavior demonstrates that changes in the LT contribution can be largely compensated by corresponding shifts in the effective Schwarzschild precession parameter $\Upsilon$.
		
		The range of slopes for the linear degeneracy direction in equation~\eqref{eq:Upschidirection} can be obtained by substituting the injected orbital parameters, BH mass, and distance, together with the adopted prior $\gamma \sim \mathcal{U}(0.9,\,1.1)$, which yields $\mathcal{P}\in[-0.056,\,0.056]$. Combining this with the adopted uniform priors on $\Upsilon$ and $\chi$ gives the corresponding intercept range $\mathcal{C}\in[0.844,\,1.156]$.
		
		\begin{figure*}
			\centering
			\includegraphics[width=\textwidth]{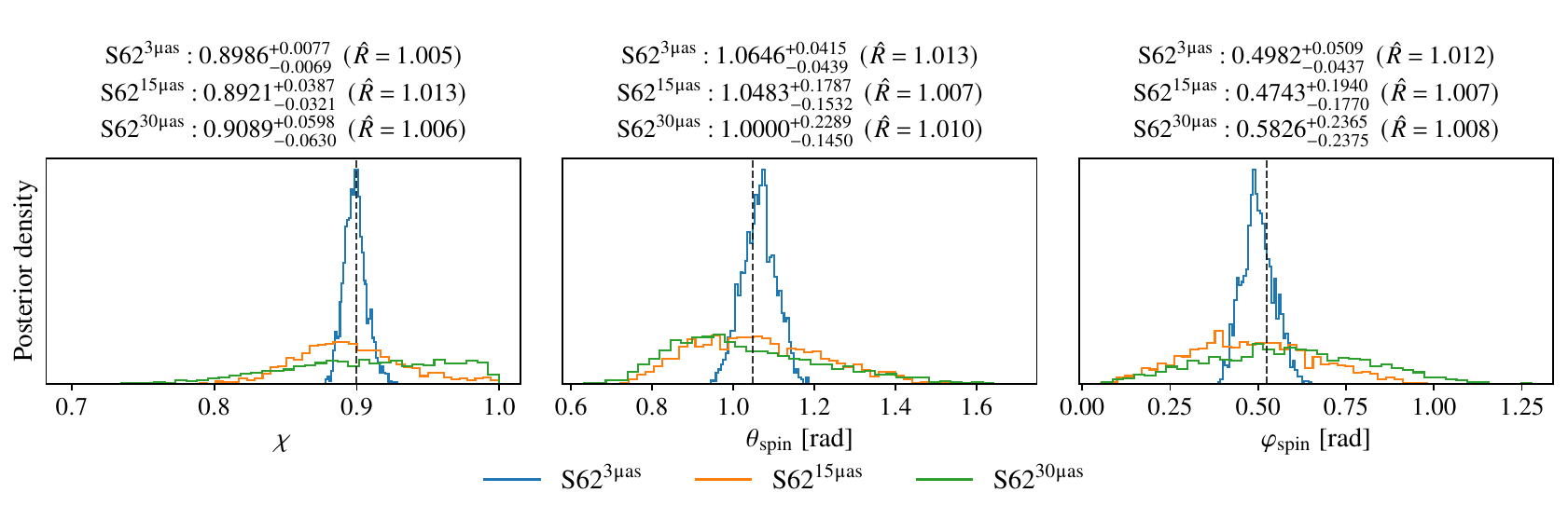}
			\vfill
			\includegraphics[width=\textwidth]{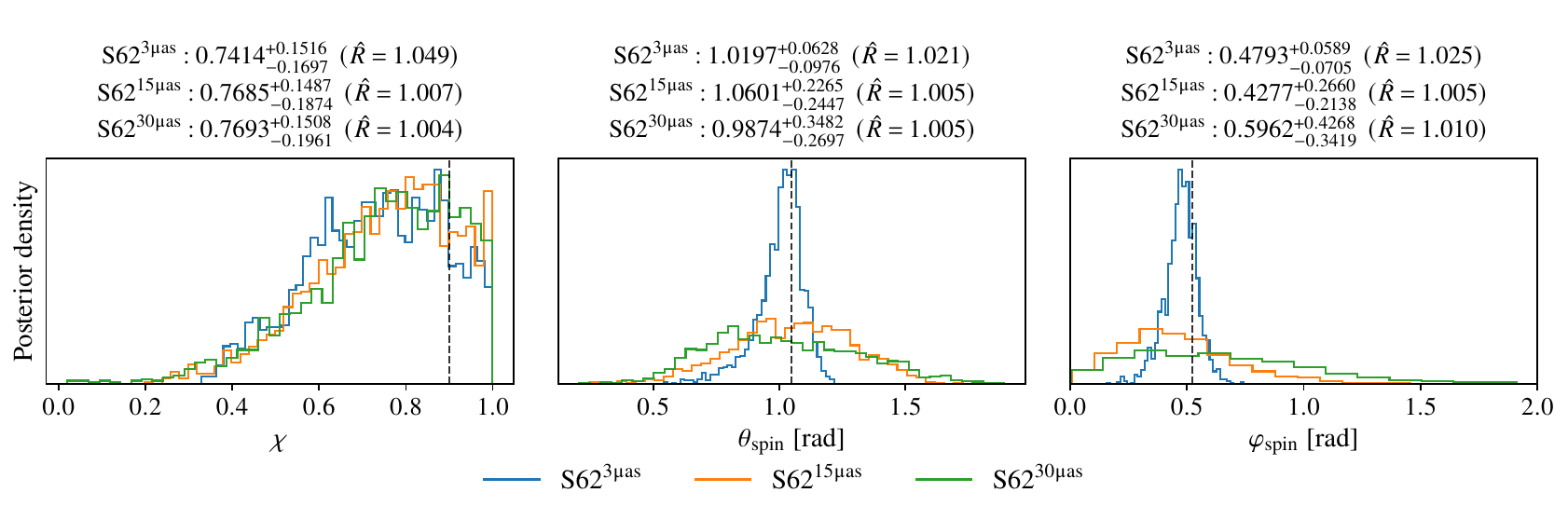}
			\caption{One-dimensional marginalized posterior distributions of the spin magnitude $\chi$ and spin orientation angles for the S62$^{3\upmu\mathrm{as}}$, S62$^{15\upmu\mathrm{as}}$ and S62$^{30\upmu\mathrm{as}}$ datasets. In the first row, the PPN parameters are fixed to their injected GR values, while in the second row they are treated as free parameters in the inference. \textit{Dashed black lines} are injected values}
			\label{fig:S62+exp}
		\end{figure*}
		\begin{figure*}
			\centering
			\includegraphics[width=0.65\columnwidth]{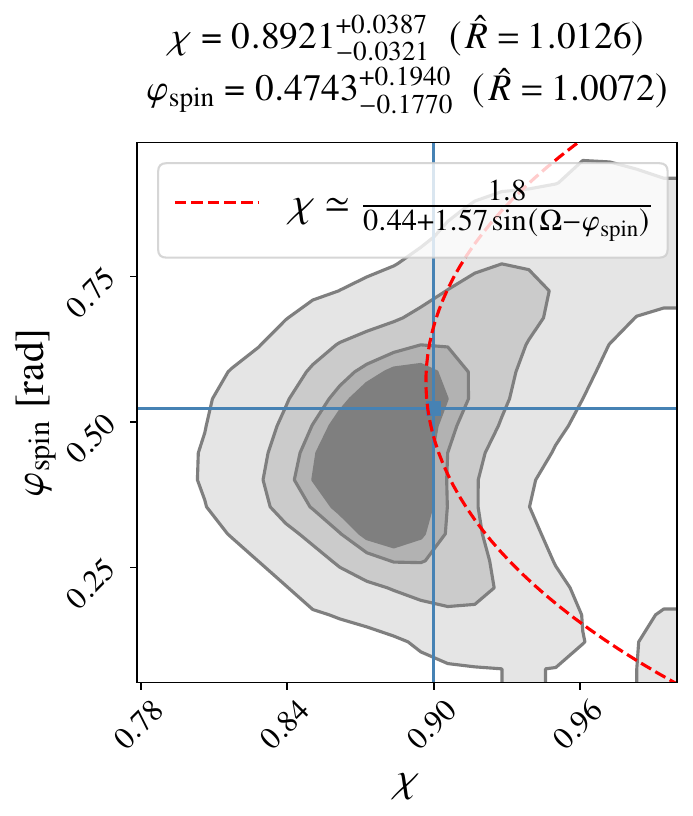}
			\includegraphics[width=0.65\columnwidth]{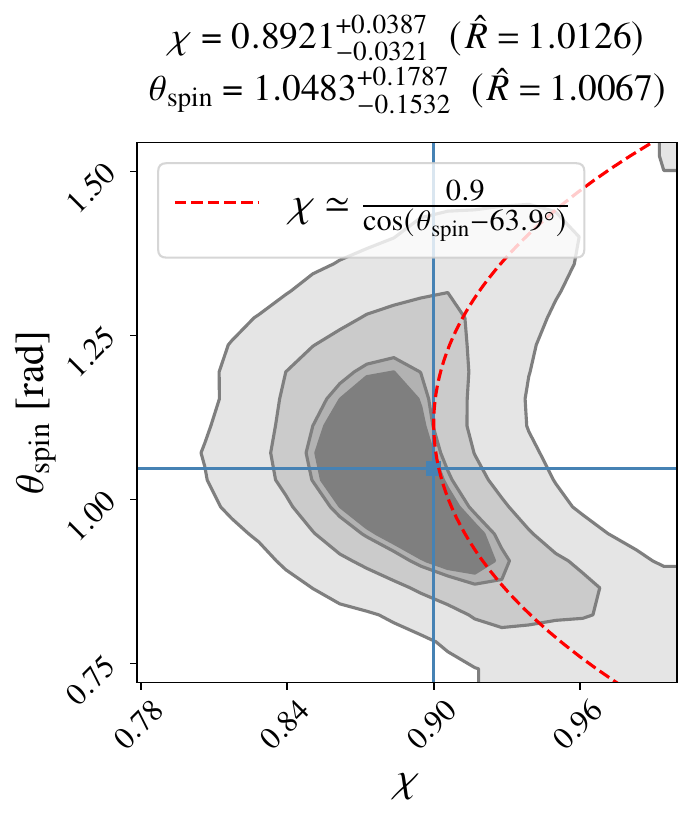}
			\caption{\textit{Left panel}: Marginalized posterior distribution in the $(\chi,\, \varphi_{\rm spin})$ plane. \textit{Right panel}: Marginalized posterior distribution in the $(\chi,\, \theta_{\rm spin})$ plane. The inference results correspond to first row of Fig.~\ref{fig:S62+exp} for the S62$^{15\upmu{\rm as}}$ dataset. The \textit{dashed red lines} indicate the approximate degeneracy directions, while the \textit{blue lines} denote the injected values.}
			\label{fig:chiangles}
		\end{figure*}
		\begin{figure*}
			\centering
			\includegraphics[width=0.7\textwidth]{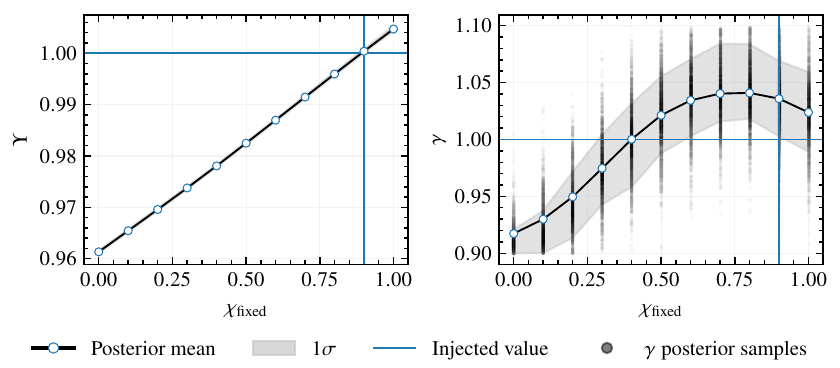}
			\caption{Posterior mean and $1\sigma$ credible intervals for the $\Upsilon$ and $\gamma$ parameters obtained by fixing the spin parameter $\chi$ in the range $[0,1]$ with increments of $0.1$, using the S62$^{3\upmu\mathrm{as}}$ dataset. The uncertainty in $\Upsilon$ remains approximately constant across the explored $\chi$ range, with $\sigma_{\Upsilon}\sim4\times10^{-4}$.}
			\label{fig:chimodels}
		\end{figure*}
		
		\begin{figure*}
			\centering
			\includegraphics[width=0.65\columnwidth]{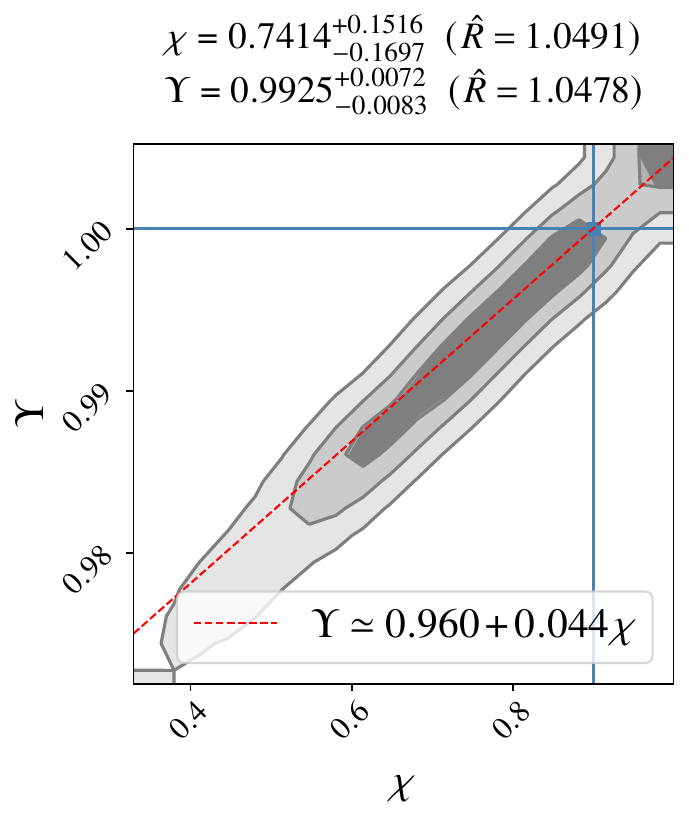}
			\includegraphics[width=0.65\columnwidth]{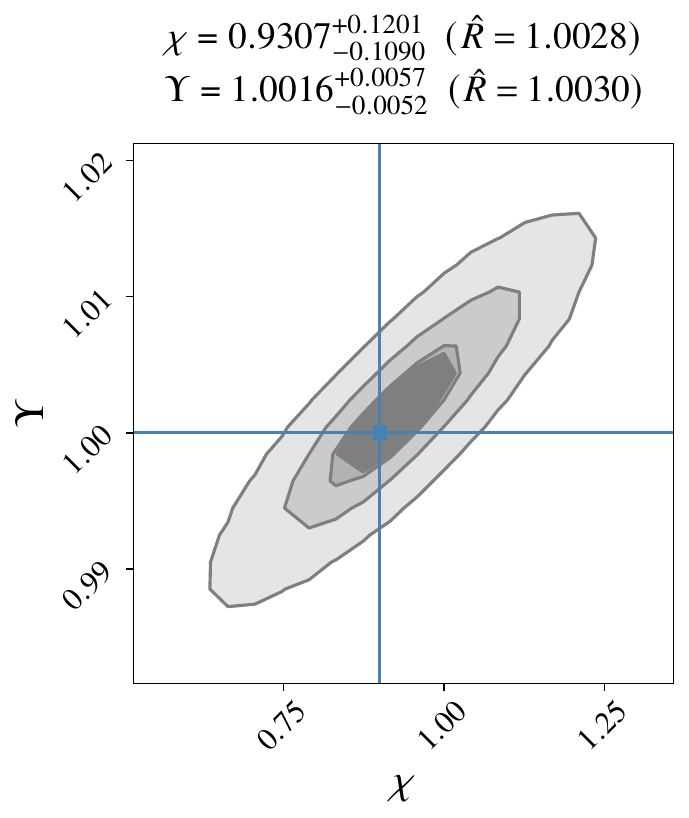}
			\caption{Marginalized posterior distribution in the $(\chi,\,\Upsilon)$ plane. \textit{Left panel}: inference result corresponding to second row of Fig.~\ref{fig:S62+exp} for the S62$^{3\upmu{\rm as}}$ dataset; the \textit{dashed red line} indicates the approximate linear degeneracy direction. \textit{Right panel}: joint inference using the S62$^{3\upmu{\rm as}}$ and $\widetilde{\mathrm{S2}}^{\rm m}$ datasets. The \textit{blue lines} indicate the injected values.}
			\label{fig:cornerPPNchi}
		\end{figure*}

		This degeneracy can be partially alleviated by adopting more informative priors on the PPN parameters, or through a joint multi-star analysis utilizing targets at different radial distances. Incorporating a more distant star, which exhibits measurable 1PN precession but negligible spin-induced effects, provides an independent constraint on $\Upsilon$, thereby helping to reduce the covariance and improve the isolation of the spin parameter $\chi$ in the joint posterior.
		
		To demonstrate this, we perform a multi-star inference using the S62$^{3\upmu\mathrm{as}}$ dataset combined with a modified $\widetilde{\mathrm{S2}}^{\rm m}$ dataset. To test whether the combined data can mitigate the degeneracy, we replace the uniform spin prior with a restrictive Gaussian prior centred away from the injected value, $\chi \sim \mathcal{N}(0.5,\,0.3^2)$, while retaining uniform priors on the PPN parameters. Incorporating $\widetilde{\mathrm{S2}}^{\rm m}$ provides an anchor for the 1PN sector, tightening the constraints to $\Upsilon = 1.001 \pm 0.005$ and $\gamma = 1.002 \pm 0.004$, while substantially reducing the $\Upsilon$--$\chi$ degeneracy (right panel of Fig.~\ref{fig:cornerPPNchi}). Consequently, the spin posterior moves away from the prior-dominated regime and recovers the injected value, $\chi = 0.93 \pm 0.12$ (right panel of Fig.~\ref{fig:cornerPPNchi}). Owing to the residual uncertainty induced by the intrinsic astrometric noise of the S62$^{3\upmu\mathrm{as}}$ measurements, the inferred posterior remains sufficiently broad that its upper tail extends beyond the physical bound $\chi = 1$ at the $2\sigma$ level, yielding a $95\%$ credible interval of $[0.72, 1.14]$.

		\section{Conclusion}\label{seq:conclusion}
		
		In this work, we investigated the degeneracies between BH spin effects and PPN parameters in the relativistic orbital dynamics of S2-like and S62-like stellar configurations around $\sgr$. Using a 1PN$+$SO Hamiltonian framework in the ADM gauge together with synthetic astrometric and radial velocity observations, we performed Bayesian parameter inference for both S2-like and S62-like mock datasets.
		
		Our analysis confirms that, for observational precisions represented by the baseline S2$^{\rm m}$ and S62$^{\rm m}$ datasets, the orbital parameters and the effective Schwarzschild precession parameter $\Upsilon$ can be recovered, while the individual PPN parameters $\gamma$ and $\beta$ remain strongly degenerate. This reflects the fact that the dominant relativistic observable, namely the Schwarzschild periapsis advance, depends primarily on the linear combination $\Upsilon=(2+2\gamma-\beta)/3$ rather than on $\gamma$ and $\beta$ separately.
		
		For the enhanced S62-like datasets with microarcsecond astrometric precision, the spin-induced LT signal becomes partially measurable. When the PPN sector is fixed to its GR values, the spin parameter can be constrained with high precision, reaching uncertainties at the level of $\sim10^{-2}$ for the S62$^{3\upmu\rm{as}}$ dataset.
		
		Allowing the PPN parameters to vary simultaneously with the BH spin introduces a strong covariance between the effective Schwarzschild precession parameter $\Upsilon$ and the spin magnitude $\chi$. The fixed-spin experiments demonstrate an approximately linear degeneracy relation of the form $\Upsilon \simeq \mathcal{P}\chi + \mathcal{C}$, where the slope depends on the orbital geometry and spin orientation. This relation captures the dominant covariance structure between the PPN and spin sectors in relativistic orbital inference.
		
		Finally, we showed that this degeneracy can be reduced through joint multi-star inference, where a wider-orbit star constrains the dominant 1PN sector while a compact relativistic orbit remains sensitive to LT frame dragging.

		\section*{Acknowledgments}
		The author is grateful to the anonymous referee(s) for their constructive comments and suggestions, which helped to improve the clarity and strengthen the content of the manuscript. This work was supported by the Higher Education and Science Committee of RA (Research project \textnumero~ 24WS-1C001). 
		
		\section*{Data Availability}
		The data underlying this article will be shared on reasonable request to the corresponding author.

		
        \bibliographystyle{plainnat}

		\bibliography{refs} 
		
		
		
		\appendix
		
		\section{Coordinate Transformations and Orbit Integration Procedure}\label{sec:app:framesIntegration}
		\subsection{Frame transformations and Newtonian initial conditions}\label{subsec:app:TransformIC}
		To integrate the equations of motions, equation~\eqref{eq:HamiltonEq}, we must specify the initial phase-space coordinates in the BH frame. 
		We choose the reference time to be the periapsis passage epoch $T_{\rm p}$ and specify the orbit using the instantaneous Keplerian elements $(a, e, \iota, \Omega, \omega)$. 
		At this instant, the Newtonian position and momentum vectors are constructed in the observer frame.
		The orbital plane is described by the orthonormal right-handed triad $(\myvecun{p}, \myvecun{q}, \myvecun{l})$, where
		$\myvecun{l} = \myvecun{p}\times \myvecun{q}$ (see Fig.~\ref{fig:RefFrames}). In the observer frame, the unit vectors are given by
		\begin{align}
			\myvecun{p}_{\rm obs} &=
			\begin{pmatrix}
				\cos\omega\cos\Omega - \sin\omega\sin\Omega\cos\iota \\
				\cos\omega\sin\Omega + \sin\omega\cos\Omega\cos\iota \\
				\sin\omega\sin\iota
			\end{pmatrix}, \\
			\myvecun{q}_{\rm obs} &=
			\begin{pmatrix}
				-\sin\omega\cos\Omega - \cos\omega\sin\Omega\cos\iota \\
				-\sin\omega\sin\Omega + \cos\omega\cos\Omega\cos\iota \\
				\cos\omega\sin\iota
			\end{pmatrix}.
		\end{align}
		At $t = T_{\rm p}$, the position and corresponding momentum vector are expressed by
		\begin{align}
			\myvec{r}_{\rm obs}(T_{\rm p}) &= r_{\rm peri}^{\rm{Newt}}\,\myvecun{p}_{\rm obs},\\
			\myvec{p}_{\rm obs}(T_{\rm p}) &= p_{\rm peri}^{\rm{Newt}}\,\myvecun{q}_{\rm obs},
		\end{align}
		where $r_{\mathrm{peri}}^{\mathrm{Newt}}=a_{\mathrm{sma}}(1 - e)$ and $p_{\rm peri}^{\rm{Newt}}=\left({G\mbh (1+e)}/{r_{\mathrm{peri}}^{\mathrm{Newt}}}\right)^{{1}/{2}}$ are determined from the standard Keplerian relations.
		The numerical integration is performed in the BH frame. The transformation of the state from the observer frame is given by
		\begin{equation}
			\left(\myvec{r}_{\rm BH}(0), \myvec{p}_{\rm BH}(0)\right)
			=
			\left(\SkytoBH\,\myvec{r}_{\rm obs}(\Tp),\;
			\SkytoBH \,\myvec{p}_{\rm obs}(\Tp)\right),
		\end{equation}
		where the rotation matrix is constructed from the BH frame basis vectors:
		\begin{equation}
			\SkytoBH =
			\left(
			\myvecun{x}_{\rm BH},\;
			\myvecun{y}_{\rm BH},\;
			\myvecun{z}_{\rm BH}
			\right)^{\mathrm T}.
		\end{equation}
		The integration is performed in geometrized units $G = c = M_{\rm BH} = 1$, after numerical integration physical units are restored by appropriate rescalings. The equations of motion are integrated both forward and backward in time from the reference point $t=0$ in the BH frame. After the numerical solution $(\myvec{r}_{\rm BH}(t),\myvec{p}_{\rm BH}(t))$ is obtained we go to the velocities from canonical momentum using the first Hamilton equation equation~\eqref{eq:HamiltonEq} and projecting to observer frame using rotation matrix $\BHtoSky\hspace{-0.8ex}\equiv \SkytoBH^{\mathrm{T}}$. Finally, the observer time is obtained from the emission time in the BH frame using equation~\eqref{eq:obstime}.
		
		\subsection{Post Newtonian initial conditions}\label{subsec:app:1PNIC}
		To ensure consistency between the initial conditions and the 1PN+SO dynamics when the periapsis is closer to the BH (especially for the S62 star case), the Newtonian periapsis construction described above is refined at the 1PN+SO level. The orbital geometry (i.e., the orientation defined by $(\myvecun{p}, \myvecun{q}, \myvecun{l})$) is preserved, while the conserved energy $E$ and angular momentum $L$ are computed from the Keplerian elements using their Newtonian relations and subsequently treated as invariants of the motion. In contrast, the radial coordinate at periapsis is no longer taken to be the Newtonian value $r_{\rm peri}^{\rm Newt}$, but is instead determined self-consistently from the 1PN+SO dynamics.
		
		The relativistic periapsis radius $r_{\rm peri}$ is then obtained by solving the energy constraint
		\begin{equation}
			E = \mathcal{H}\left(r_{\rm peri}\,\myvecun{p},\tfrac{L}{r_{\rm peri}}\,\myvecun{q}\right),
		\end{equation}
		
		This equation is solved numerically for $r_{\rm peri}$ using a root-finding procedure, with the Newtonian periapsis $r_{\rm peri}^{\rm Newt}$ serving as an initial guess. This guarantees that the initial phase-space point lies exactly on the relativistic energy surface, avoiding spurious radial oscillations at the start of the integration. The resulting expressions generalize the Newtonian construction by retaining the same conserved quantities $(\mathcal{H}, \myvec{L}\cdot\myvecun{s}, |\myvec{L}|)$ while replacing the Keplerian periapsis distance with its post-Newtonian counterpart.
		
		For completeness, we note that at the 1PN level one may alternatively employ a quasi-Keplerian parametrization of the orbit. However, this introduces additional gauge-dependent orbital elements \citep{damour1985general,memmesheimer2004third} and is not required for the present approach.

		
		\subsection{Time mapping during inference}\label{subsec:app:timemapping}
		During inference we have access only to the observed times $t_{\rm{obs}}$, while the corresponding emission times $t_{\rm{em}}$ are not directly known but are required to evaluate the dynamical evolution and relativistic corrections. This mapping can be obtained in several ways. For instance, one may solve the implicit equation~\eqref{eq:obstime} numerically at each observation epoch, or employ approximate inversion schemes as commonly used in S-star analyses \citep{abuter2018detection2,heissel2022dark}. 
		Alternatively, instead of explicitly inverting , one may construct a dense forward solution for the orbital evolution over a buffered time interval and evaluate the corresponding observables on a finely sampled grid, then interpolated onto the observed timestamps during inference. But this can be computational not efficient, and also not accurate in periapsis passage points.
		
		Instead, we adopt a different strategy. During inference, the observed time $t_{\rm obs}$ is treated as the independent integration variable, while the emission time $t_{\rm em}$ is promoted to a dynamical variable. The physical system is originally defined in terms of the state $\myvec{y} = (\myvec{r}, \myvec{p})$ and evolves with respect to emission time as
		\begin{equation}
			\frac{d\myvec{y}}{dt_{\rm em}} = \myvec{F}(\myvec{y}),
		\end{equation}
		where $\myvec{F}(\myvec{y})$ denotes the Hamiltonian vector field. This system can be reparameterized using
		\begin{equation}
			\frac{dt_{\rm em}}{dt_{\rm obs}} = \left(1 + \frac{d\Delta_{\rm delay}}{dt_{\rm em}}\right)^{-1},
		\end{equation}
		where $\Delta_{\rm delay} := \Delta_{\rm R} + \Delta_{\rm Sh}$.
		
		With this transformation, the augmented system evolved with respect to $t_{\rm obs}$ becomes
		\begin{align}
			\frac{d\myvec{y}}{dt_{\rm obs}} &= 
			\frac{\myvec{F}(\myvec{y})}{1 + \frac{d\Delta_{\rm delay}}{dt_{\rm em}}}, \\
			\frac{dt_{\rm em}}{dt_{\rm obs}} &= 
			\frac{1}{1 + \frac{d\Delta_{\rm delay}}{dt_{\rm em}}}.
		\end{align}
		The derivative $d\Delta_{\rm delay}/dt_{\rm em}$ can be computed analytically or evaluated efficiently using a Jacobian vector product.
		The solver therefore integrates the augmented state $\tilde{\myvec{y}} = (\myvec{r}, \myvec{p}, t_{\rm em})$ directly as a function of $t_{\rm obs}$, eliminating the need for explicit inversion of $t_{\rm em}(t_{\rm obs})$ or delay-offset corrections.
		
		We use the {\tt{Diffrax}}\footnote{\url{https://docs.kidger.site/diffrax}} JAX-based library \citep{kidger2021on} with 8-th order Runge--Kutta method \citep{prince1981high}
		
		\section{Gravitational lensing model}\label{sec:app:GL}
		In this section, we describe the weak-field GL correction applied to the astrometric position. The correction is applied
		as a post-processing step after the orbital integration and after transforming the trajectory to the observer frame. We include only the displacement of the primary image, which corresponds to the observed stellar image, while the
		secondary and higher-order images are neglected.
		
		The unlensed angular position on the sky and the lensed primary-image position are denoted by
		\begin{equation}
			\myvec{\theta}_0:=
			\begin{pmatrix}
				\Delta\alpha\\
				\Delta\delta
			\end{pmatrix},
			\qquad
			\myvec{\theta}_{\rm GL}:=
			\begin{pmatrix}
				\Delta\alpha_{\rm GL}\\
				\Delta\delta_{\rm GL}
			\end{pmatrix},
		\end{equation}
		The lensing displacement is then
		\begin{equation}
			\updelta\myvec{\theta}_{\rm GL}
			=
			\myvec{\theta}_{\rm GL}-\myvec{\theta}_0
			=
			\begin{pmatrix}
				\updelta\alpha_{\rm GL}\\
				\updelta\delta_{\rm GL}
			\end{pmatrix}.
		\end{equation}
		
		For reference, in the usual weak-field thin-lens approximation the point-mass lens equation gives two images. The angular separation of the primary image is (see for example \citep{wambsganss1998gravitational,narayan1996lectures})
		\begin{equation}
			\theta_{+}=
			\frac{1}{2}
			\left[
			\theta_0+\sqrt{\theta_0^2+4\theta_{\rm E}^2}
			\right],
		\end{equation}
		where the PPN-corrected Einstein angle is \citep{cao2017test}
		\begin{equation}
			\theta_{\rm E}
			=
			\left(\frac{1+\gamma}{2}\right)^{1/2}
			\left(
			2r_{\rm s}
			\frac{D_{\rm ls}}{D_{\rm l}D_{\rm s}}
			\right)^{1/2}.
		\end{equation}
		Here \(D_{\rm l}\), \(D_{\rm s}\), and \(D_{\rm ls}\) are the observer--lens, observer--source, and lens--source distances, respectively. Since \(\myvecun{z}_{\rm obs}\) points from the BH towards the observer, a source
		behind the BH satisfies \(\myvecun{z}_{\rm obs}\cdot\myvec{r}<0\). Thus, to leading order in \(r/\Dobs\),
		\begin{equation}
			D_{\rm l}=\Dobs,
			\qquad
			D_{\rm s}\simeq
			\Dobs-\myvecun{z}_{\rm obs}\cdot\myvec{r},
			\qquad
			D_{\rm ls}\simeq
			\max\left(-\myvecun{z}_{\rm obs}\cdot\myvec{r},0\right).
		\end{equation}
		The thin-lens expression is useful as a reference, but it is not adopted for the final astrometric correction below.
		
		Instead, we use the finite-distance weak-field result derived for S-stars by \citet{bozza2012observing}. We define the alignment angle \(\psi\) by 
		\begin{equation}
			\psi=\arccos\left(
			-\frac{\myvecun{z}_{\rm obs}\cdot\myvec{r}}{r}
			\right),
		\end{equation}
		where \(r=|\myvec{r}|\). With this definition, \(\psi=0\) corresponds to a source located exactly behind the BH, while \(\psi=\pi\) corresponds to a source located between the BH and the observer.
		
		Starting from the weak-field lens equation and taking the Galactic centre limit \(\Dobs\gg r\), \citet{bozza2012observing} obtain for the primary-image astrometric shift
		\begin{equation}
			\Delta\theta_{\rm GL}^{\rm GR}
			=
			\frac{r_{\rm s}}{\Dobs}
			\frac{\cos^3(\psi/2)}{\sin(\psi/2)} .
		\end{equation}
		The corresponding small-angle thin-lens expression is recovered only for
		\(\psi\ll1\),
		\begin{equation}
			\Delta\theta_{\rm GL}^{\rm thin}
			\simeq
			\frac{2r_{\rm s}}{\Dobs\,\psi}.
		\end{equation}
		As discussed by \citet{bozza2012observing}, this small-angle expression already differs from the finite-distance result at the level of order ten percent for \(\psi\simeq30^\circ\). We therefore use the finite-distance expression rather
		than the small-angle thin-lens approximation.
		The lensing displacement used in this work is therefore
		\begin{equation}\label{eq:GLgamma}
			\Delta\theta_{\rm GL}
			=
			\left(\frac{1+\gamma}{2}\right)
			\frac{r_{\rm s}}{\Dobs}
			\frac{\cos^3(\psi/2)}{\sin(\psi/2)} .
		\end{equation}
		The displacement of the primary image is radial on the sky, away from the BH.
		Therefore,
		\begin{equation}
			\updelta\myvec{\theta}_{\rm GL}
			=
			\Delta\theta_{\rm GL}
			\frac{\myvec{\theta}_0}{\theta_0}.
		\end{equation}
		or in component form
		\begin{equation}
			\updelta\alpha_{\rm GL}
			=
			\Delta\theta_{\rm GL}
			\frac{\Delta\alpha}{\theta_0},
			\qquad
			\updelta\delta_{\rm GL}
			=
			\Delta\theta_{\rm GL}
			\frac{\Delta\delta}{\theta_0}.
		\end{equation}
		
		The total astrometric impact of gravitational lensing is quantified by $\left|\updelta\myvec{\theta}_{\rm GL}\right|\equiv \Delta\theta_{\rm GL}$
		
		We do not include the spin dependent lensing of the photon trajectory in this correction. The BH spin is already included in the orbital dynamics through the SO term in the Hamiltonian, and therefore affects the apparent position through the integrated stellar trajectory. The spin contribution to the light propagation is a separate higher order correction to the weak field lensing map. We therefore neglect photon spin-lensing in the present
		postprocessing and keep only the leading PPN mass-deflection contribution. The
		time delay model is kept as described in Section~\ref{subsec:obsevables}.
		
		\section{Time delay astrometric contributions}\label{sec:app:DelayContribut}
		The apparent astrometric contributions associated with the R\o mer and Shapiro delays are defined operationally by comparing astrometric positions evaluated at the same observed time. These quantities are not direct angular deflections of the photon trajectory. Instead, they measure the apparent displacement caused by the fact that propagation-time effects change the mapping between emission time and observed time.
		
		First, we define the R\o mer delay contribution. We compare two unlensed astrometric positions: one evaluated without any propagation-time delay, and one including only the R\o mer delay. Denoting these astrometric coordinates by
		$\Delta\alpha_{\rm E}(t_{\rm obs})$, $\Delta\delta_{\rm E}(t_{\rm obs})$ and $\Delta\alpha_{\rm R}(t_{\rm obs})$, $\Delta\delta_{\rm R}(t_{\rm obs})$,
		respectively, we define
		\begin{align}
			\updelta\alpha_{\rm R}(t_{\rm obs})
			&=
			\Delta\alpha_{\rm R}(t_{\rm obs})
			-
			\Delta\alpha_{\rm E}(t_{\rm obs}),
			\\
			\updelta\delta_{\rm R}(t_{\rm obs})
			&=
			\Delta\delta_{\rm R}(t_{\rm obs})
			-
			\Delta\delta_{\rm E}(t_{\rm obs}).
		\end{align}
		The corresponding total R\o mer delay astrometric contribution is
		\begin{equation}
			\Delta\theta_{\rm R}(t_{\rm obs})=	
			\left[
			\updelta\alpha_{\rm R}^2(t_{\rm obs})
			+
			\updelta\delta_{\rm R}^2(t_{\rm obs})
			\right]^{\frac{1}{2}}.
		\end{equation}
		
		The Shapiro delay contribution is defined in an analogous way. We compare two unlensed astrometric positions evaluated at the same observed time: one including only the R\o mer delay, and one including both the R\o mer and Shapiro delays. Denoting these astrometric coordinates by $\Delta\alpha_{\rm R}(t_{\rm obs})$, $\Delta\delta_{\rm R}(t_{\rm obs})$ and $\Delta\alpha_{\rm R+Sh}(t_{\rm obs})$, $\Delta\delta_{\rm R+Sh}(t_{\rm obs})$, respectively, we define
		\begin{align}
			\updelta\alpha_{\rm Sh}(t_{\rm obs})
			&=
			\Delta\alpha_{\rm R+Sh}(t_{\rm obs})
			-
			\Delta\alpha_{\rm R}(t_{\rm obs}),
			\\
			\updelta\delta_{\rm Sh}(t_{\rm obs})
			&=
			\Delta\delta_{\rm R+Sh}(t_{\rm obs})
			-
			\Delta\delta_{\rm R}(t_{\rm obs}).
		\end{align}
		The corresponding total Shapiro delay astrometric contribution is
		\begin{equation}
			\Delta\theta_{\rm Sh}(t_{\rm obs})=
			\left[
			\updelta\alpha_{\rm Sh}^2(t_{\rm obs})
			+
			\updelta\delta_{\rm Sh}^2(t_{\rm obs})
			\right]^{\frac{1}{2}}.
		\end{equation}
		
		In practice, the comparisons above are performed after interpolating the corresponding astrometric positions onto a common observed time grid. This ensures that the differences isolate the apparent astrometric shifts caused by the time remapping, rather than by evaluating the orbit at different observed epochs.
		
		\section{Orbital Precessions}\label{sec:app:precession}
		First of all lets write equation of motion in accleration form
		\begin{align}
			\myvec{a}=  -\frac{G\mbh}{r^3}\myvec{r} + \myvec{a}_{\rm{1PN}} + \myvec{a}_{\rm{1.5PN}}
		\end{align}
		where
		\begin{align}\label{eq:dotomegaContributions}
			\myvec{a}_{\rm{1PN}}&= \frac{1}{c^2}\frac{G\mbh}{r^3}\left[\left(\frac{2G\mbh}{r}(\gamma+\beta)-\gamma v^2\right)\myvec{r}+2(\gamma +1)(\myvec{r}\cdot\myvec{v})\myvec{v}\right],\\
			\myvec{a}_{\rm{1.5PN}}&=-\frac{1}{c^3}\frac{G^2\mbh^2}{r^3}\chi (\gamma + 1)\left[2\myvec{v}\times\myvecun{s}-3\dot{r}\frac{\myvec{r}\times \myvecun{s}}{r}-3\frac{(\myvec{r}\times\myvec{v})\cdot\myvecun{s}}{r^2}\myvec{r}\right]
		\end{align}
		which can deriwed straightforvard from equation~\eqref{eq:HamiltonEq} (see also \citep{2013degn.book.M,poisson2014gravity}). For deriving periapsis advance per orbit we can use the Laplace-Runge–Lenz vector
		\begin{equation}
			\myvec{e}=\frac{L}{G\mbh}\myvec{v}\times \myvecun{l}-\frac{\myvec{r}}{r}
		\end{equation}
		which is constant of motion in Keplerian case, then
		\begin{equation}\label{eq:dotomega}
			\dot{\omega}=\frac{\dot{\myvec{e}}\cdot \myvecun{q}}{e}-\dot{\Omega}\cos\iota,
		\end{equation}
		where the first term represents the intrinsic rotation of the Laplace-Runge–Lenz vector within the instantaneous orbital plane, and the second term is the geometric contribution arising from the motion of the ascending node $\Omega$ as the orbital plane precesses in space. After differentiation and seting $\myvec{a}$ the $\myvec{r}/{r}$ part cancels out with Newtonian acceleration, the 1PN and 1.5PN give the following contribution in equation~\eqref{eq:dotomega}
		\begin{align}
			\dot{\omega}_{\rm 1PN}
			=&\frac{1}{c^2}\frac{L}{er^3}\myvecun{l}\cdot\myvecun{q}\times\Bigg[\left(\frac{2G\mbh}{r}(\gamma + \beta)-\gamma v^2\right)\myvec{r}+4(\gamma+1)(\myvec{r}\cdot\myvec{v})\myvec{v}\Bigg]\nonumber\\
		\end{align}
		\begin{align}
			\dot{\omega}_{\rm 1.5PN}
			=&-\frac{1}{c^3}\frac{G\mbh}{er^3}\chi (\gamma + 1) \Bigg[ L(\myvec{v}\cdot\myvecun{q})(\myvecun{s}\cdot\myvecun{l})+\myvecun{s}\cdot(\myvecun{q}\times\myvec{v})(\myvec{r}\cdot\myvec{v})\Bigg]\nonumber\\
			&-\dot{\Omega}_{\rm{1.5PN}}\cos\iota
		\end{align}
		1PN term have no contribution to the nodal precession. To extract the 1.5PN contributions to the nodal precession rate $\dot{\Omega}$ and the inclination evolution $\dot{\iota}$, it is convenient to introduce the orthonormal triad $(\myvecun{n}, \myvecun{m}, \myvecun{l})$. The unit vector $\myvecun{n}$ points along the line of nodes, while $\myvecun{m}$ is orthogonal to both $\myvecun{n}$ and $\myvecun{l}$. In the observer (inertial) frame, these vectors take the form
		\begin{align}
			&\myvecun{n}_{\rm{obs}}= \begin{pmatrix}
				\cos\Omega \\
				\sin\Omega\\
				0
			\end{pmatrix}, 
			\myvecun{m}_{\rm{obs}} = \begin{pmatrix}
				-\cos\iota\sin\Omega \\
				\cos\iota \cos\Omega\\
				\sin\iota
			\end{pmatrix},
			\myvecun{l}_{\rm{obs}}= \begin{pmatrix}
				\sin\iota\sin\Omega \\
				-\sin\iota\cos\Omega\\
				\cos\iota
			\end{pmatrix}
		\end{align}
		The time evolution of $\myvecun{l}$ can be decomposed as
		\begin{align}
			\dot{\myvecun{l}} 
			= -\dot{\iota}\,\myvecun{m} 
			+ \dot{\Omega}\,\sin\iota \,\myvecun{n}.
		\end{align}
		In other hand from dynamics we have a
		\begin{equation}
			\dot{\myvecun{l}}=\Omega_{\rm{LT}}\,(\myvecun{s}\times\myvecun{l}),
		\end{equation}
		and therefore 
		\begin{align}
			&\dot{\iota}=\Omega_{\rm{LT}}\,(\myvecun{s}\cdot\myvecun{n}),\quad	\dot{\Omega}\sin\iota=\Omega_{\rm{LT}}\,(\myvecun{s}\cdot\myvecun{m})\\
			&\Omega_{\rm{LT}} :=\frac{G^2\mbh^2}{c^3 r^3}\chi(\gamma + 1)
		\end{align}
		Only 1.5PN part have contribution to $\dot{\iota}$ and $\dot{\Omega}$. To obtain secular (orbit-averaged) variations, we average over one radial period, which can be written as
		\begin{align}
			\Delta\omega_{\rm{1PN}}   &= \left( \frac{2+2\gamma-\beta}{3} \right)6\pi\epsilon,\\
			\Delta\omega_{\rm{1.5PN}} &= -2\pi\epsilon^{\frac{3}{2}}\chi (\gamma + 1)\left[2(\myvecun{s}\cdot\myvecun{l})+(\myvecun{s}\cdot\myvecun{m})\cot\iota\right],\\
			\sin\iota\Delta\Omega_{\rm{1.5PN}}&=2\pi\epsilon^{\frac{3}{2}}\chi (\gamma + 1)\,(\myvecun{s}\cdot\myvecun{m}),\\
			\Delta\iota_{\rm{1.5PN}}&=2\pi\epsilon^{\frac{3}{2}}\chi (\gamma + 1)\,(\myvecun{s}\cdot\myvecun{n}),
		\end{align}
		where $\epsilon = {G\mbh}/{a_{\rm{sma}}(1-e^2)c^2}$. The spin-induced periapsis precession $\Delta\omega_{\rm 1.5PN}$ for S62-like orbital elements, presented in Table~\ref{tab:dataparams}, evaluated over the full range of spin orientation angles, $\theta_{\rm spin}\in[0,\pi]$ and $\varphi_{\rm spin}\in[0,2\pi)$, is shown in Fig.~\ref{fig:precrv}. The figure demonstrates the degeneracy associated with the spin orientation angles, while the corresponding radial velocity at periapsis passage provides additional information that can help alleviate this degeneracy. The figure also includes numerical estimates of $\Delta\omega_{\rm 1.5PN}$ obtained from full three-dimensional orbital integrations, rather than from the projected orbit approximation. The numerical precessions are extracted directly from the evolution of the osculating orbital elements over successive periapsis passages.
		
		\begin{figure}
			\includegraphics[width=\columnwidth]{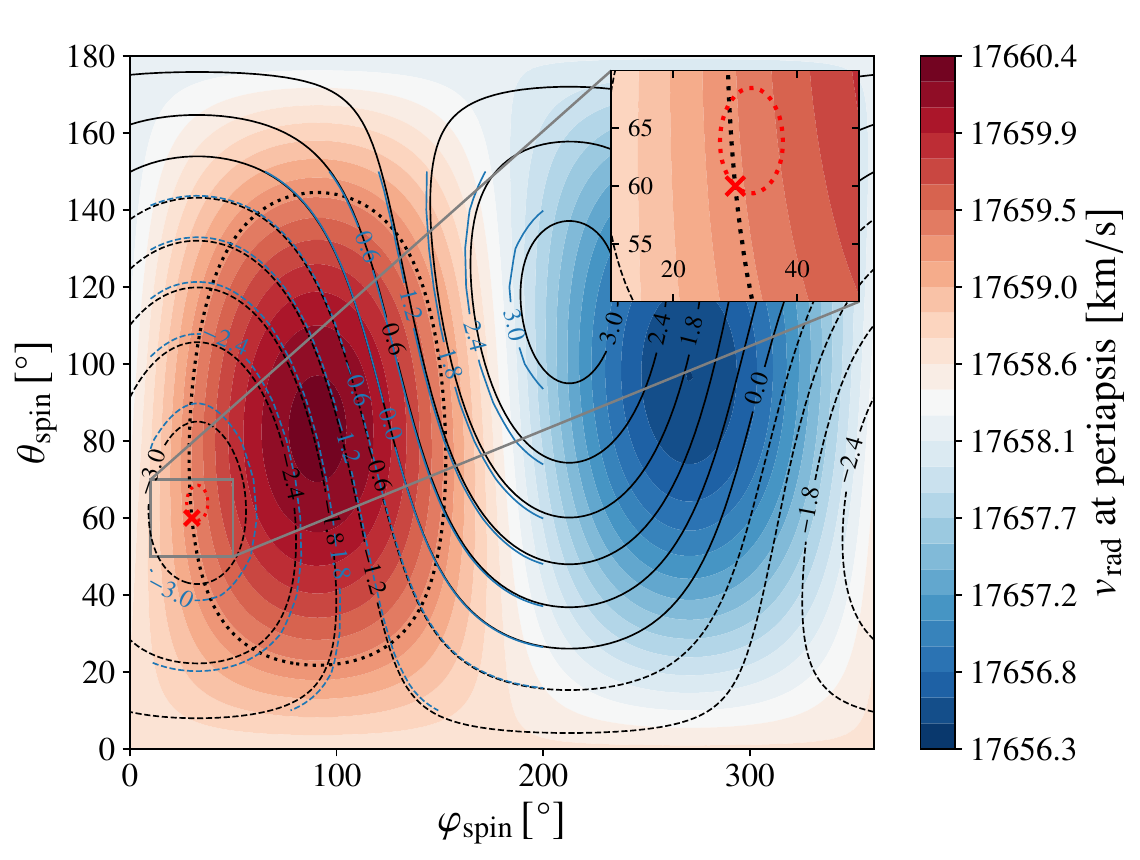}
			\caption{Spin-induced periapsis precession, $\Delta\omega_{\mathrm{1.5PN}}$, for an S62-like orbit with $\chi=0.9$ and $\epsilon \approx 1.2\times10^{-3}$, shown as a function of the spin-orientation angles $(\varphi_{\mathrm{spin}},\theta_{\mathrm{spin}})$. The corresponding radial velocity at periapsis passage is also displayed, illustrating its potential to partially reduce the correlation between the spin orientation angles. Numerical results obtained from full three-dimensional 1PN+SO orbital integrations (\textit{blue lines}) are shown for comparison with the analytical results (\textit{black lines}). The precessions are given in $\mathrm{arcmin}$ per orbit. The \textit{red cross marks} the reference configuration $(30^{\circ},60^{\circ})$, while the \textit{red} and \textit{black dotted lines} denote the $\Delta\omega_{\mathrm{1.5PN}}$ and $v_{\rm rad}$ contours passing through this reference point, respectively.}
			\label{fig:precrv}
		\end{figure}
		
		\section{MAP details}\label{sec:app:MAPDet}
		We perform MAP estimation of the orbital and relativistic parameters by minimizing the log posterior probability equation~\eqref{eq:LMAP}. Given a set of model parameters $\Theta$, the forward model predicts:
		\begin{equation}
			\hat{\Delta \alpha}(t), \quad
			\hat{\Delta \delta}(t), \quad
			\hat{v}_{\rm rad}(t).
		\end{equation}
		We define normalized residuals as:
		\begin{align}
			r_{\alpha,i} &= 
			\frac{\hat{\Delta \alpha}(t_i) - \Delta \alpha_{\rm obs}(t_i)}
			{\sigma_{\alpha,i}}, \\
			r_{\delta,i} &= 
			\frac{\hat{\Delta \delta}(t_i) - \Delta \delta_{\rm obs}(t_i)}
			{\sigma_{\delta,i}}, \\
			r_{{\rm RV},j} &= 
			\frac{\hat{v}_{\rm rad}(t_j) - v_{\rm rad,obs}(t_j)}
			{\sigma_{{\rm RV},j}}.
		\end{align}
		Assuming independent Gaussian uncertainties, we define the log-likelihood as
		\begin{equation}
			\ln \widetilde{\mathcal{L}}
			=
			-\frac{1}{2}
			\left[
			\sum_i r_{\alpha,i}^2
			+
			\sum_i r_{\delta,i}^2
			+
			\sum_j r_{{\rm RV},j}^2
			\right].
		\end{equation}
		The corresponding MAP objective is globally tempered by the factor $1/N_{\rm tot}$, as defined in equation~\eqref{eq:LMAP}.
		This expression differs from the standard Gaussian log-likelihood by an overall normalization factor $1/N_{\rm tot}$. Such a global rescaling does not affect the location of the MAP estimate, as it corresponds to multiplying the log-likelihood by a positive constant. However, it rescales the curvature of the posterior and would therefore affect uncertainty estimates if used for full Bayesian inference. In our analysis, this normalized form is employed only during MAP optimization for numerical stability, while the standard (unscaled) Gaussian log-likelihood is used for MCMC sampling and uncertainty quantification.
		
		For all model parameters, we adopt priors that regularize the inference while avoiding strong biases. The specific choices are summarized in Table~\ref{tab:MAPparametrization}. For scale parameters, we use log‑normal priors, which naturally enforce positivity and reflect the typical order‑of‑magnitude uncertainty in such quantities. For parameters with physically allowed ranges (e.g., eccentricity, angles), we employ uniform priors. To maintain differentiability and improve optimization stability, these uniform priors are implemented as soft quadratic penalties outside the allowed interval, rather than as hard cutoffs. This approach prevents the optimizer from stepping into forbidden regions while still providing a well‑defined gradient.
		
		The initial values of all parameters except $\Tp$ are randomly drawn from their priors: within the allowed range for uniform priors, and within $2\sigma$ for normal and log‑normal distributions. The initial value of $\Tp$ is fixed and set to the first time at which the observed radial velocity $v_{\rm rad,obs}(t_j)$ reaches its maximum. The initial learning rate, the learning rate scheduler, and the number of steps for the Adam optimizer are chosen manually and depend on the initial parameter values. We note that the minimization procedure exhibits some dependence on the initialization. In particular, for the S62$^{\rm m}$ data, the result is sensitive to the initial value of $a$. Furthermore, the obtained solution is not necessarily the global minimum of $\mathcal{L}_{\rm MAP}$. Therefore, we perform 100 random initializations in parallel for both the S2$^{\rm m}$ and S62$^{\rm m}$ datasets and adopt the parameter set corresponding to the lowest value of $\mathcal{L}_{\rm MAP}$. Fig.~\ref{fig:MAPloss} shows the evolution of the loss for three different starting points, for both stars using mock data.
		\begin{figure}[H]
			\includegraphics[width=\columnwidth]{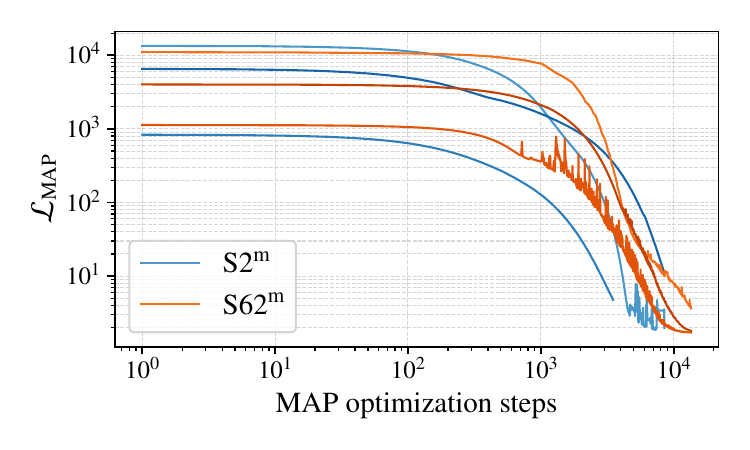}
			\caption{Evaluation of the MAP loss for the S2$^{\rm m}$ and S62$^{\rm m}$ datasets using three different initializations for each dataset.}
			\label{fig:MAPloss}
		\end{figure}
		We use {\tt{Optax}}\footnote{\url{https://github.com/google-deepmind/optax}} library for JAX compilable Adam optimizer.

		\section{MCMC posterior distributions and summary tables}\label{sec:app:mcmcplots}
		We generate the corner plots using the \citep{corner} package.
		\begin{figure*}
			\centering
			\centering
			\includegraphics[width=\textwidth]{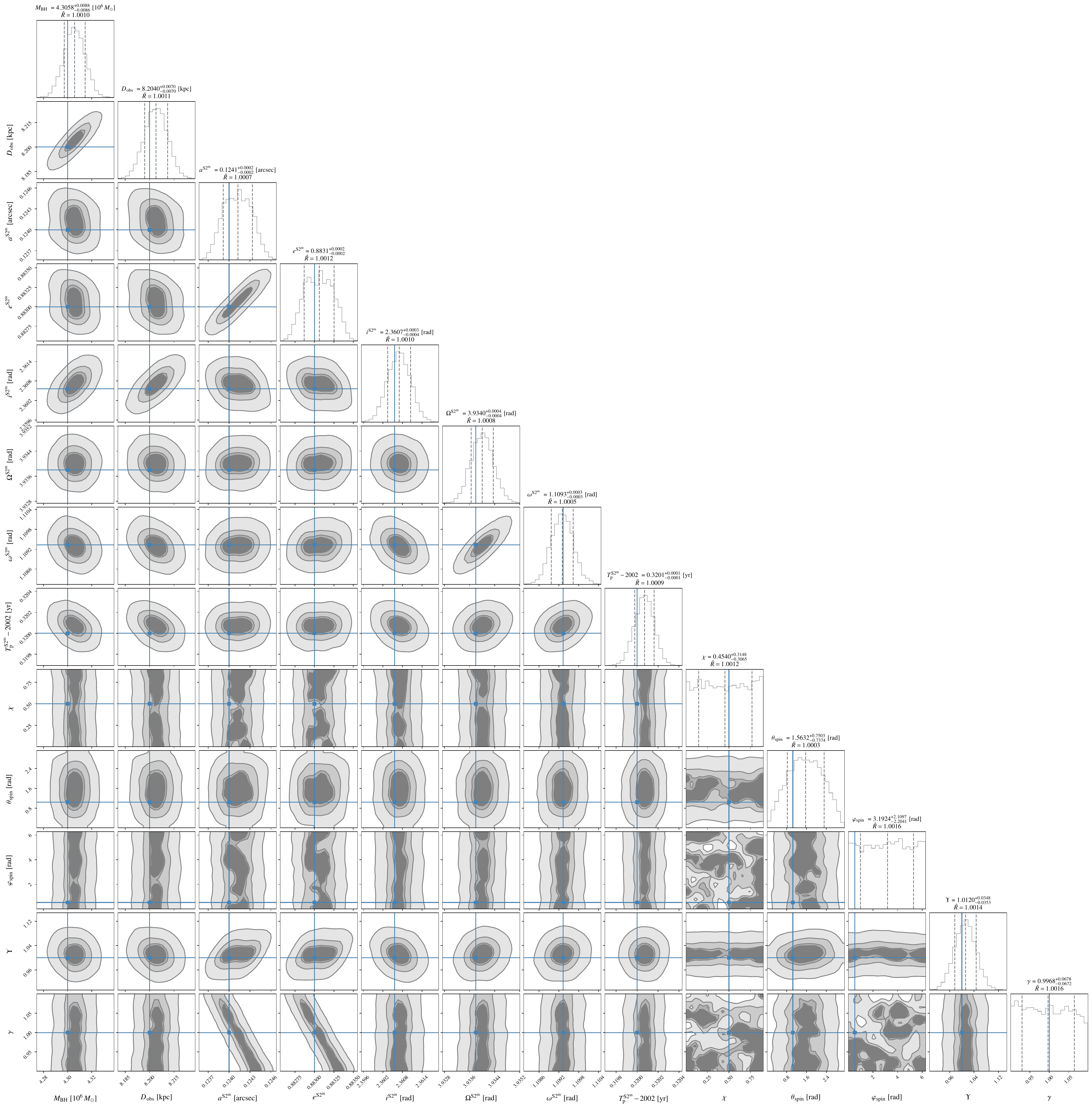}
			\caption{Corner plot showing the one- and two-dimensional marginalized posterior distributions of the parameter set $\Theta$, inferred from fitting the mock S2$^{\rm m}$ dataset. The \textit{dashed lines} in the one-dimensional posteriors indicate the 16th, 50th, and 84th percentiles, while the contours denote enclosed posterior probability regions of 30\%, 40\%, 68\%, and 95\%. The \textit{blue lines} and \textit{blue markers} indicate the injected parameter values.}
			\label{fig:S2corner}
		\end{figure*}
		\begin{figure*}
			\centering
			\includegraphics[width=\textwidth]{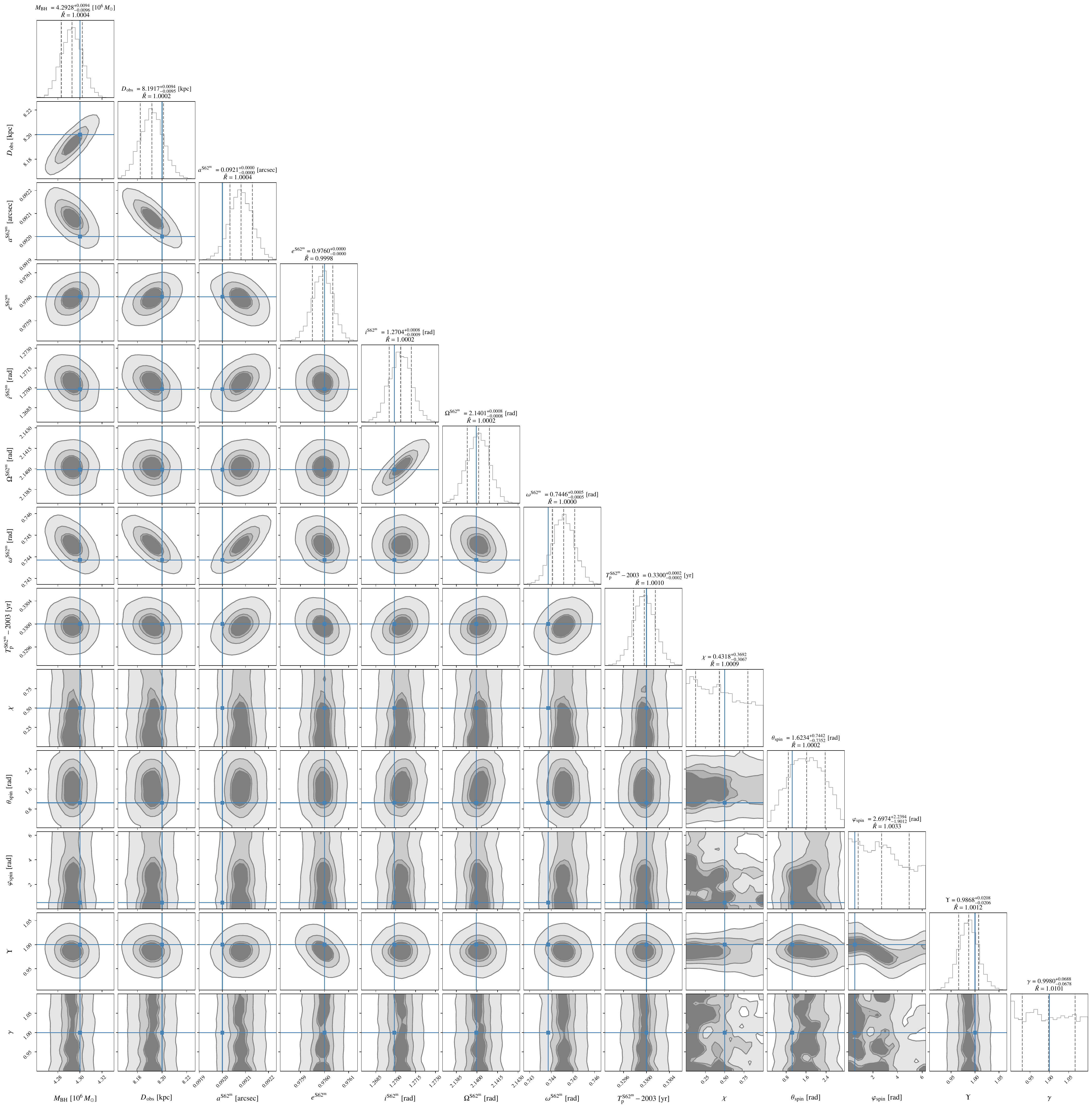}
			\caption{Corner plot showing the one- and two-dimensional marginalized posterior distributions of the parameters $\Theta$, inferred from fitting a mock dataset of the S62$^{\rm m}$ star. The \textit{dashed lines} in the one-dimensional posteriors indicate the 16th, 50th, and 84th percentiles, while the contours denote enclosed posterior probability regions of 30\%, 40\%, 68\%, and 95\%. The \textit{blue lines} and \textit{blue markers} indicate the injected parameter values.}
			\label{fig:S62corner}
		\end{figure*}
		\begin{figure*}
			\centering
			\includegraphics[width=0.65\columnwidth]{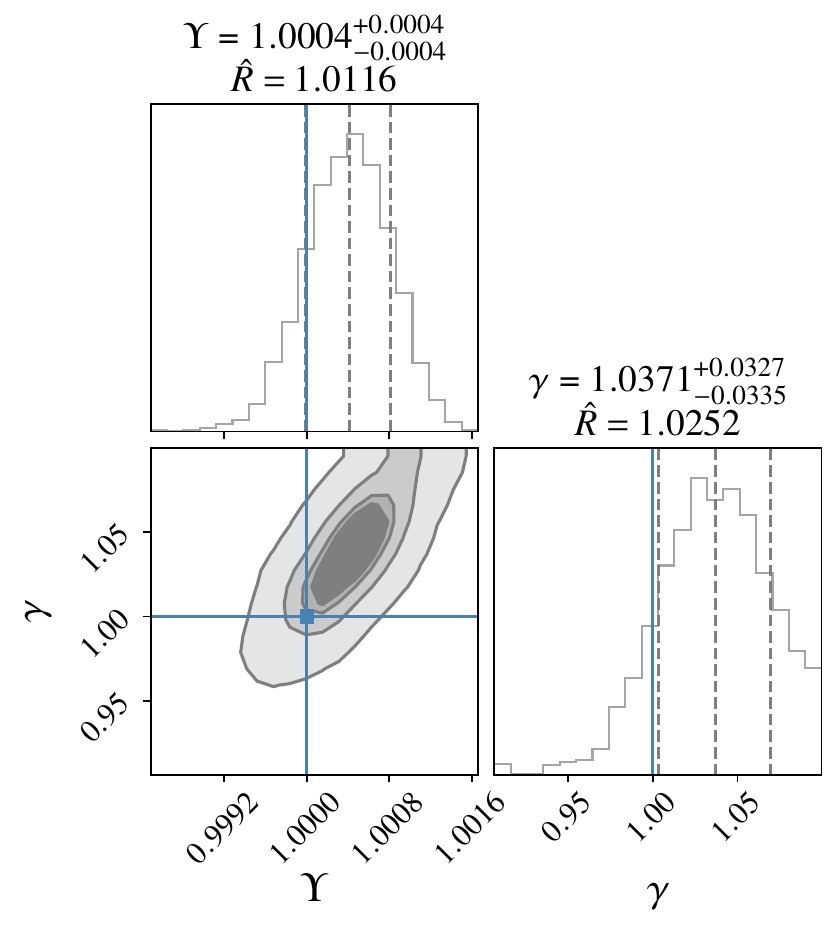}
			\includegraphics[width=0.65\columnwidth]{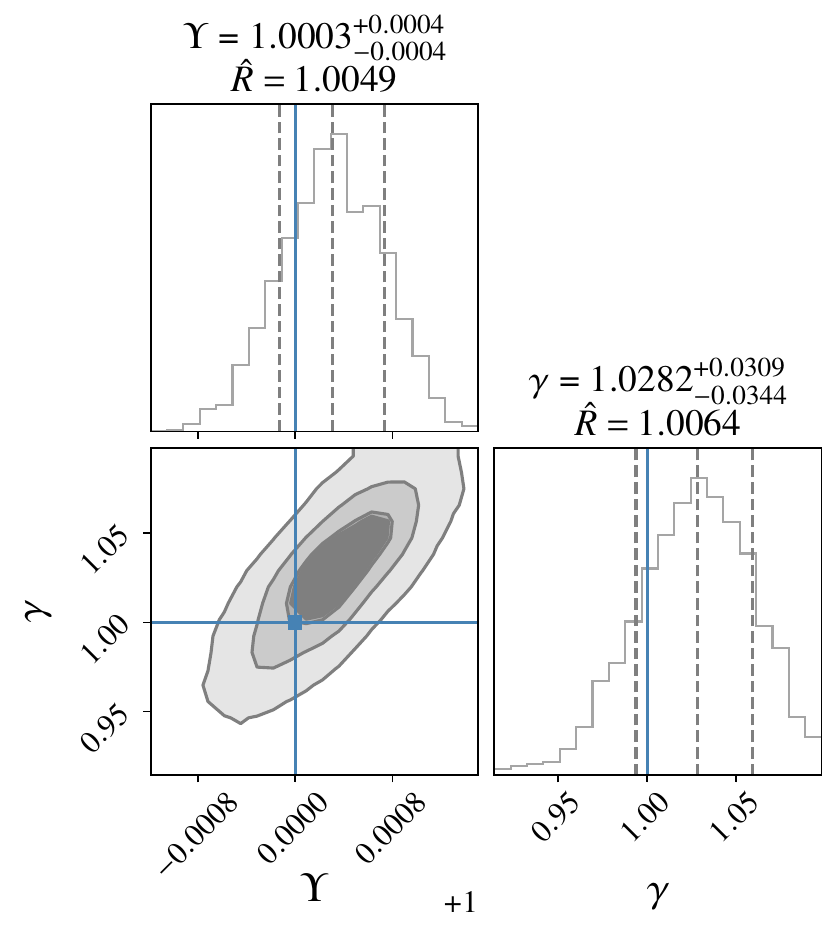}
			\caption{Marginalized posterior distribution in the $(\Upsilon,\gamma)$ plane. \textit{Left panel}: Inference result for the S62$^{3\upmu\rm{as}}$ dataset, corresponding to the $\chi_{\rm fixed}=0.9$ value of Fig.~\ref{fig:chimodels}. \textit{Right panel}: Inference result for the S62$^{3\upmu\rm{as}}_{\rm GL}$ dataset when $\chi$ fixed to injected value.}
			\label{fig:Gammagamma}
		\end{figure*}
		
		\begin{table*}[h!]
			\centering
			\caption{Orbital parameter uncertainties for different astrometric precisions for S62. The top panel shows the results obtained with the PPN parameters fixed to their GR values (corresponding to the first row of Fig.~\ref{fig:S62+exp}), while the bottom panel shows the results with the PPN parameters treated as free parameters (corresponding to the second row of Fig.~\ref{fig:S62+exp}). For all cases, the inferred mean values remain within the $1\sigma$--$2\sigma$ confidence regions of the injected values.}
			\label{tab:orbitalsigma}
			\begin{tabular}{lccccc}
				\hline
				& $\sigma_{a}\, [\rm arcsec]$ & $\sigma_{e}$ & $\sigma_{i} \,[\rm rad]$ 
				& $\sigma_{\Omega} \,[\rm rad]$ & $\sigma_{\omega} \,[\rm rad]$ \\
				\hline
				\multicolumn{6}{l}{\textbf{PPN parameters fixed}} \\
				S62$^{30\mu\rm as}$  
				& $2.00\times10^{-6}$ 
				& $1.00\times10^{-5}$  
				& $2.50\times10^{-5}$  
				& $2.80\times10^{-5}$ 
				& $3.60\times10^{-5}$ \\
				
				S62$^{15\mu\rm as}$  
				& $1.00\times10^{-6}$   
				& $5.00\times10^{-6}$  
				& $1.20\times10^{-5}$  
				& $1.30\times10^{-5}$ 
				& $1.70\times10^{-5}$ \\
				
				S62$^{3\mu\rm as}$   
				& $1.82\times10^{-7}$   
				& $1.07\times10^{-6}$  
				& $2.70\times10^{-6}$  
				& $3.01\times10^{-6}$ 
				& $3.57\times10^{-6}$ \\
				
				
				\multicolumn{6}{l}{\textbf{PPN parameters free}} \\
				S62$^{30\mu\rm as}$  
				& $2.00\times10^{-6}$ 
				& $1.00\times10^{-5}$  
				& $1.28\times10^{-4}$  
				& $1.40\times10^{-4}$ 
				& $5.70\times10^{-5}$ \\
				
				S62$^{15\mu\rm as}$  
				& $1.00\times10^{-6}$   
				& $5.00\times10^{-6}$  
				& $6.80\times10^{-5}$  
				& $7.60\times10^{-5}$ 
				& $2.90\times10^{-5}$ \\
				
				S62$^{3\mu\rm as}$   
				& $4.26\times10^{-7}$   
				& $1.10\times10^{-6}$  
				& $1.49\times10^{-5}$  
				& $1.68\times10^{-5}$ 
				& $6.34\times10^{-6}$ \\
				\hline
			\end{tabular}
		\end{table*}


\end{document}